\documentclass[pdflatex,sn-mathphys-num]{sn-jnl}

\usepackage{amssymb, amsmath, amsfonts}
\usepackage{verbatim, color, epsfig}
\usepackage{bm}
\usepackage{booktabs}
\usepackage{multirow}
\usepackage{placeins}
\usepackage{hyperref}
\usepackage[compatibility=false]{caption}
\usepackage{subcaption}
\usepackage{amsthm}
\theoremstyle{plain}

\theoremstyle{definition}

\theoremstyle{remark}

\begin{document}

\title[Tail Estimation via Sample Splitting]{{Tail} Estimation via Sample Splitting: A Two-Sample Framework for Stable-like Distributions}

\author*[1,2]{\fnm{Cornelis J.} \sur{Potgieter}}\email{c.potgieter@tcu.edu}
\author[2]{\fnm{Jacques} \sur{van Appel}}
\author[3]{\fnm{Sudharshan} \sur{Samaratunga}}

\affil*[1]{\orgdiv{Department of Mathematics}, \orgname{Texas Christian University},
  \orgaddress{\city{Fort Worth}, \state{TX}, \country{USA}}}

\affil[2]{\orgdiv{Department of Statistics}, \orgname{University of Johannesburg},
  \orgaddress{\city{Auckland Park}, \country{South Africa}}}

\affil[3]{\orgdiv{Department of Statistics and Data Science}, 
  \orgname{Southern Methodist University},
  \orgaddress{\city{Dallas}, \state{TX}, \country{USA}}}

\abstract{
Stable distributions provide a flexible framework for modeling heavy-tailed and skewed data, with the stability index $\alpha$ quantifying tail heaviness. We propose a new semiparametric approach that leverages the two-sum closure property of stable distributions within a location-scale framework, \textcolor{black}{where the induced scale parameter satisfies $\sigma = 2^{1/\alpha}$}. The method constructs two pseudo-independent samples from a single observed sample via repeated random splitting and empirical convolution, and then estimates $\sigma$ using weighted least squares applied to empirical quantiles. We also propose a bootstrap procedure that uses a reduced number of sample splits together with extrapolation to estimate standard errors, and we demonstrate good finite-sample performance of this procedure. This approach avoids intractable likelihood calculations and offers computational advantages over maximum likelihood estimation. We establish consistency and asymptotic properties of the estimator and assess its finite-sample performance through simulation studies. Beyond the benefit of substantial computational savings, results indicate competitive accuracy, particularly in heavy-tailed settings.
}


\maketitle

\section{Introduction}

In many disciplines, ranging from finance and insurance to telecommunications, physics, and environmental science (see, e.g., \citet{mcculloch1996}, \citet{wang2023}, \citet{genolini2017} and \citet{botai2011}), data frequently display behavior inconsistent with finite-variance models such as the normal distribution. Heavy-tailed distributions provide a more flexible framework for modeling such phenomena, as their ability to capture extreme outcomes better reflects the underlying stochastic mechanisms giving rise to the data. Accurate parameter estimation is thus crucial, not only for theoretical development but also for practical applications involving risk assessment, anomaly detection, and robust forecasting. 

Among the families of heavy-tailed distributions, the stable distribution stands out due to its foundational role in generalizing the central limit theorem, see \citet{gnedenko1954}, and its intrinsic connection to self-similar processes and fractal behavior (see, e.g., \citet{feller1991}). Introduced by Paul  L\'{e}vy in 1924, the stable distribution arises naturally in the study of sums of independent and identically distributed (\textit{i.i.d.}) random variables. A non-degenerate random variable $X$ is said to have a stable distribution if and only if, for all $n\geq2$, there exist constants $a_n\in\mathbb R$ and $b_n>0$ such that
\begin{align}
    X_1+X_2+\ldots+X_n\stackrel{d}{=}a_n+b_nX.
\end{align}
Here, $X_1,X_2,\ldots,X_n$ are independent copies of $X$ and the symbol $\stackrel{d}{=}$ is used to denote equality in distribution. The random variable $X$ is called strictly stable if and only if $a_n=0$ for all values of $n$. For a recent and more comprehensive overview of the existing results on stable distributions, see \citet{nolan2020}.

The stable distribution is typically defined in terms of its characteristic function. A random variable $X$ is said to have a stable distribution  $\mathrm{S}(\alpha,\beta,\gamma,\delta)$ if, for all real $t$, its characteristic function is given by
\begin{equation}
    \label{eq:cfk0}
    \psi_X(t) = 
    \begin{cases}
        \exp{\bigl[-\gamma^\alpha |t|^\alpha\bigl\{1+i\beta\tan{\left(\frac{\pi\alpha}{2}\right)}\operatorname{sign}t\bigl(|\gamma t|^{1-\alpha}-1\bigr)\bigr\}+i\delta t\bigr]}  &    \hspace{-0.15cm}\text{if $\alpha\neq1$,}  \\[1mm]
        \exp{\bigl[-\gamma|t|\bigl\{1+\frac{2i\beta}{\pi}\operatorname{sign}t\log{\bigl(\gamma|t|\bigr)}\bigr\}+i\delta t\bigr]}  & \hspace{-0.15cm}\text{if $\alpha=1$,}
    \end{cases}
\end{equation}
where $\alpha\in(0,2]$ is the stability index, $\beta\in[-1,1]$ is a skewness parameter, $\gamma>0$ is a scale parameter, and $\delta\in\mathbb{R}$ is a location parameter. The parameter $\alpha$, which governs the tail heaviness of the distribution, is often the primary object of inferential interest. As $\alpha$ decreases, the tails of the distribution become heavier. For $\alpha\leq 1$, the mean does not exist, while for $1<\alpha<2$, the variance is also undefined. In general, stable distributions possess finite moments of order $p$ only for $p<\alpha$. That is, $\mathbb{E}\bigl(|X|^p\bigr)<\infty$ for $p<\alpha$.

A defining feature of stable distributions is their closure under convolution: sums of \textit{i.i.d.} stable random variables remain stable with the same characteristic exponent $\alpha$. In practice, stable distributions are an attractive option for modeling data that exhibit heavy tails and skewness; these features can be easily captured by a stable distribution. However, stable distributions generally lack closed-form density functions. The three special cases in which a closed-form expression exists for the density function are the normal, Cauchy and L\'{e}vy distributions. When setting $\alpha=2$,  a Normal$\big(\delta,2\gamma^2\big)$ distribution results. Similarly, setting $(\alpha,\beta)=(1,0)$ yields a Cauchy$(\gamma,\delta)$ distribution, while $(\alpha,\beta)=(1/2,1)$ results in a L\'{e}vy distribution with $\gamma$ and $\delta+\gamma$ the scale and location parameters, respectively.  Crucially, all non-degenerate instances of the stable distribution are continuous with infinitely differentiable densities. 

Given the intractability of the likelihood and the absence of finite higher-order moments, estimation of stable parameters has proven challenging. Existing parameter estimation methods fall broadly into three categories: maximum likelihood estimation (MLE), quantile-based methods, and characteristic function methods. Quantile-based techniques include both classical sample quantiles and estimators based on extreme order statistics. While some applications require estimating all four parameters, in many instances the parameter of greatest interest is the stability index $\alpha$, which determines the tail heaviness of the distribution. The focus of this paper is on the estimation of the tail parameter $\alpha$. 

\citet{dumouchel1973,dumouchel1975} did early work on the MLE, relying on data binning to approximate the likelihood function. \citet{MiDoCh99} used a fast Fourier transform to estimate the parameters, while \citet{nolan2001} developed numerical quadrature routines. \citet{lombardi2007} and \citet{buckle1995} developed classic Bayesian methodology, while \citet{peters2012} proposed a likelihood-free Bayesian approach.  However, even with modern tools, directly maximizing the stable likelihood remains computationally intensive and numerically delicate, limiting its practical adoption.

\citet{famaroll1968,famaroll1971} developed estimators of $\alpha$ using order statistics, but their method only applied to symmetric stable distributions for $\alpha \in [1,2]$. \citet{mcculloch1986} developed the quantile method for both symmetric and skewed stable distributions and $\alpha \in [0.5,2]$, now popular in applications. The Hill estimator, see \citet{hill1975}, is a popular measure of tail heaviness for distributions with Pareto-like tails, and can also be adapted to estimate $\alpha$. However, the Hill estimator tends to have large bias in small to moderately sized samples. Authors that have considered estimators making direct use of the characteristic function include \citet{press1972}, \citet{paulson1975}, \citet{koutr1980} and \citet{brockwell1981}.

In this paper, we propose a new estimator for the tail index that combines a semiparametric perspective with practical computational advantages. \textcolor{black}{The approach exploits the location-scale structure of the stable laws, where the sum of two independent copies of a stable random variable differs from the original by a location shift and a scale factor $\sigma = 2^{1/\alpha}$. Our procedure estimates this scale parameter $\sigma$ directly.} The method involves repeatedly creating two groupings of the underlying observed data: one retaining the original (observed) scale, and the other forming all possible pairwise sums of observations. This ``split-sample'' construction enables us to exploit two-sample scale estimation techniques in the present context.

We adapt the quantile-based perspective of \citet{hsieh1995}, also considered by \citet{potgieter2012}, which was originally developed for comparing two samples from the same location-scale family, to the present setting. While the stable distribution motivates the method, the proposed estimator does not assume a fully parametric model and can be viewed as operating within a \textit{semiparametric framework}. That is, although the stable distribution is a special case, the estimation method is not tied to a specific parametric likelihood. This flexibility, combined with the estimator’s reliance only on order statistics, results in a procedure that is simpler to implement and computationally more efficient than full likelihood methods, particularly given the current limitations of stable likelihood maximization routines.

The remainder of the paper is organized as follows. Section~\ref{sec:location-scale} reviews the location-scale representation of stable distributions and makes explicit the connection between the scale parameter $\sigma$ in the two-sample setting and the tail index $\alpha$. Section~\ref{sec: Sample Spliting} formally defines the split-sample estimator and adapts the two-sample quantile method of Potgieter and Lombard to the current framework for estimating $\sigma$. Theoretical justification is also provided. Section~\ref{sec: Simulations} presents simulation results that offer empirical support for the split-sample approach, demonstrating favorable root mean squared error (RMSE) performance relative to existing methods. Section~\ref{sec: Real-Data} provides an illustration using real data, and Section~\ref{sec: Conclusion} concludes with a summary and discussion of future directions.

\section{Location-Scale Model} \label{sec:location-scale}

Stable distributions exhibit a distinctive structure: the sum of any two \textit{i.i.d.} stable random variables can be expressed as a location-scale transformation of another variable with the same distribution. This property, which we refer to as \emph{two-sum closure}, provides a natural motivation for working within a location-scale framework, where the scale parameter is directly linked to the stability index $\alpha$.

While two-sum closure holds exactly for stable distributions, it is not unique to them. As noted by \citet[Problem~VI.13.3]{feller1991}, this property is necessary but not sufficient for stability, suggesting the existence of a broader class of distributions that satisfy this type of identity. However, this class remains poorly characterized and appears to include only a small number of known examples beyond the stable case; to our knowledge, only a single explicit example has been identified in the literature.

In this section, we adopt the location-scale model not only as a characterization of stable distributions, but also as a semiparametric framework that supports robust, quantile-based estimation without requiring full specification of a likelihood. Let $X, X',$ and $X''$ be \textit{i.i.d.} random variables, and define $Y = X' + X''$. Suppose these random variables satisfy
\begin{equation}
    \label{eq:loc_scale}
    Y \stackrel{d}{=}\mu+\sigma X,
\end{equation}
where $\mu \in \mathbb{R}$ and $\sigma > 0$ are constants.  Then, $X$ and $Y$ belong to the same location-scale family of distributions. Moreover, in the stable case, the scale parameter $\sigma$  of \eqref{eq:loc_scale} relates directly to the stability index $\alpha$ via $\sigma = 2^{1/\alpha}$. Thus, estimating $\alpha$ is equivalent to estimating the scale parameter $\sigma$ under this location-scale representation. This equivalence forms the basis of the estimation strategy developed in the remainder of the paper.

To define the two-sample estimator, assume we have two independent samples, $X_1, \ldots, X_n$ and $Y_1, \ldots, Y_m$, drawn from distributions $F$ and $G$, respectively, satisfying \eqref{eq:loc_scale}. Our aim is to estimate the scale parameter $\sigma$. While this semiparametric problem has received considerable attention in the literature, our goal here is not to implement an ``optimal'' estimator, but rather to prioritize computational efficiency. Readers interested in more refined approaches may consider the work of \citet{potgieter2012}, who propose a nonparametric method known as asymptotic likelihood (AL), and \citet{potgieter2016nonparametric}, who address the problem using empirical characteristic functions (ECFs), building on ideas from \citet{hall2013new}, who developed a testing procedure for the location-scale family null hypothesis. However, the approach we adopt here is most closely aligned with that of \citet{hsieh1995}.

Let $F$ and $G$ (and correspondingly $f$ and $g$) denote the distribution (density) functions of the random variables $X$ and $Y$, respectively. Furthermore, let $\hat{F}$ and $\hat{G}$ denote the usual empirical distribution functions (EDFs) based on independent samples $X_1, X_2, \ldots, X_n$ and $Y_1, Y_2, \ldots, Y_m$, drawn \textit{i.i.d.} from $F$ and $G$, respectively. %Specifically, the EDF for $X$ is given by
%\begin{align}
%    \hat{F}(x) = \frac{1}{n} \sum_{i=1}^{n} \mathbb{I}(X_i \leq x),
%\end{align}
%where $\mathbb{I}(\cdot)$ denotes the indicator function, which equals 1 if the condition inside holds and 0 otherwise. Consequently, for $x < X_{(1)}$, $\hat{F}(x) = 0$, and at each order statistic, the EDF satisfies
%\begin{align}
%    \hat{F}\{X_{(j)}\} = \frac{j}{n},\quad j = 1, 2, \ldots, n.
%\end{align}
%Similarly, $\hat{G}$ is defined based on the order statistics of $Y$. 
The corresponding empirical quantile functions, $\hat{Q}_F = \hat{F}^{-1}$ and $\hat{Q}_G = \hat{G}^{-1}$, are defined in the usual way as the left-continuous inverses of $\hat{F}$ and $\hat{G}$, satisfying 
\begin{align}
\hat{F}\bigl\{\hat{Q}_F(t)\bigr\} \geq t \quad \text{and} \quad \hat{G}\bigl\{\hat{Q}_G(t)\bigr\} \geq t.
\end{align}
When the location-scale model holds, the quantile functions $F^{-1}$ and $G^{-1}$ satisfy the relation
\begin{equation}
    \label{quantile_relation}
    G^{-1}(t) = \mu + \sigma F^{-1}(t), \quad 0 < t < 1,
\end{equation}
which implies that estimating $\sigma$ may be treated as a regression problem between the empirical quantiles of $Y$ and $X$. Formally, let $\mathbf{t}_{k} = [t_1, \ldots, t_k]^\top$ be a vector of quantile levels with $0 < t_1 < \cdots < t_k < 1$ and define $\hat{\mathbf{q}}_F = \hat{Q}_F(\mathbf{t}_k)$ and $\hat{\mathbf{q}}_G = \hat{Q}_G(\mathbf{t}_k)$ as the vectors of empirical quantiles evaluated at these levels. \textcolor{black}{Following \citet{hsieh1995}, the number of quantile levels $k$ may increase with sample size, albeit at a relatively slow asymptotic rate. In finite samples, however, the choice of $k$ can affect performance, particularly for heavy-tailed distributions.} Given a choice of $\mathbf{t}_k$, we estimate $\sigma$ by performing a weighted least squares (WLS) regression of $\hat{\mathbf{q}}_G$ on $\hat{\mathbf{q}}_F$.

To construct appropriate regression weights, we utilize the asymptotic covariance structure of the empirical quantiles. Under standard regularity conditions (including the existence of continuous, positive densities at the relevant quantiles), this covariance is given by
\begin{equation}
    \label{covariance}
    \mathbb{C}\mathrm{ov}\bigl\{\hat{Q}_F(t_j),\hat{Q}_F(t_k)\bigr\} = \frac{1}{n}\frac{\min(t_j, t_k) - t_j t_k }{f\bigl\{F^{-1}(t_j)\bigr\}f\bigl\{F^{-1}(t_k)\bigr\}} + \mathrm{o}\big(n^{-1}\big)
\end{equation}
with an analogous expression holding for $\hat{Q}_G$. These results follow from the functional delta method applied to the empirical quantile process; see Corollary~21.5 in \citet{van2000}.

Because the $X$ and $Y$ samples are independent, the residuals $\varepsilon_j$ in the regression model
\begin{equation}
    \hat{Q}_G(t_j) = \mu + \sigma \hat{Q}_F(t_j) + \varepsilon_j,   
\end{equation}
have asymptotic covariance structure
%\begin{align}
%    \label{cov_residual}
%    &\mathbb{C}\mathrm{ov}\bigl\{\hat{Q}_G(t_j) - \mu - \sigma \hat{Q}_F(t_j),\hat{Q}_G(t_k) - \mu - \sigma \hat{Q}_F(t_k)\bigr\}\notag \\
%    &\qquad\qquad = \mathbb{C}\mathrm{ov}\bigl\{\hat{Q}_G(t_j), \hat{Q}_G(t_k)\bigr\} + \sigma^2 \, \mathbb{C}\mathrm{ov}\bigl\{\hat{Q}_F(t_j), \hat{Q}_F(t_k)\bigr\}  \notag \\
%    &\qquad\qquad = \frac{1}{m} \frac{\min(t_j, t_k) - t_j t_k }{ g\bigl\{G^{-1}(t_j)\bigr\} g\bigl\{G^{-1}(t_k)\bigr\}} + \frac{\sigma^2}{n}\frac{\min(t_j, t_k) - t_j t_k }{f\bigl\{F^{-1}(t_j)\bigr\}f\bigl\{F^{-1}(t_k)\bigr\}} + \mathrm{o}( n^{-1}+m^{-1}) \notag \\
%    &\qquad\qquad = \frac{n+m}{nm} \frac{\min(t_j, t_k) - t_j t_k}{g\bigl\{G^{-1}(t_j)\bigr\} g\bigl\{G^{-1}(t_k)\bigr\}} + \mathrm{o}(n^{-1}+m^{-1}).
%\end{align}
\begin{equation}
    \label{cov_residual}
    \mathbb{C}\mathrm{ov}(\varepsilon_j, \varepsilon_k) = \frac{n+m}{nm} \frac{\min(t_j, t_k) - t_j t_k}{g\bigl\{G^{-1}(t_j)\bigr\} g\bigl\{G^{-1}(t_k)\bigr\}} + \mathrm{o}\big(n^{-1}+m^{-1}\big).
\end{equation}
Assuming $m/n \rightarrow c$, this motivates using covariance matrix $\bm{\Sigma}$ with elements
\begin{equation}
    \label{eq:WLS_covariance}
    \bm{\Sigma}_{jk} = \frac{\min(t_j, t_k) - t_j t_k }{ g\bigl\{G^{-1}(t_j)\bigr\} g\bigl\{G^{-1}(t_k)\bigr\}}
\end{equation}
with the corresponding WLS weight matrix $\mathbf{W} = \bm{\Sigma}^{-1}$. \textcolor{black}{The density of the underlying distribution therefore enters through the covariance structure of the empirical quantiles, determining their relative weighting in the WLS procedure. In particular, quantiles with greater sampling variability, often those farther into the tails, receive correspondingly less weight.}

In practice, the density $g$ is unknown and must be estimated. \textcolor{black}{For very heavy-tailed distributions, direct kernel density estimation on the original scale can be numerically unstable, and standard bandwidth-selection procedures may perform poorly because of the extreme dispersion of the observations. We therefore first transform the data to a more regular scale, estimate the density on that scale, and then transform the estimate back to the original scale.} Specifically, let $M = m(Y)$ with $m(y) = \operatorname{asinh}(y)$. Then the density transformation formula gives
\begin{equation}
    \label{eq:kde}
    \hat{g}(y) = \frac{\hat{f}_M\bigl\{\operatorname{asinh}(y)\bigr\}}{\sqrt{1+y^2}},
\end{equation}
where $\hat{f}_M$ is the kernel density estimate of $\text{asinh}(Y)$, using a plug-in bandwidth. For implementation, we use the \texttt{ks} package in \texttt{R} \citep{duong2022package}; see \citet{silverman1982algorithm} and \citet{wand1994kernel} for details.

To summarize, define the $k \times 2$ design matrix $\mathbf{Q}_F = [\mathbf{1}_k \ \hat{\mathbf{q}}_F]$, where $\mathbf{1}_k$ is a $k$-vector of ones, and construct $\widehat{\bm{\Sigma}}$ with entries
\begin{align}
\widehat{\bm{\Sigma}}_{jk} = \frac{ \min(t_j, t_k) - t_j t_k }{\hat{g}\bigl\{\hat{Q}_G(t_j)\bigr\} \, \hat{g}\bigl\{\hat{Q}_G(t_k)\bigl\}}.
\end{align}
The WLS estimator of the model parameters is 
\begin{align}
    [\hat{\mu},\hat{\sigma}]^\top = \big[\mathbf{Q}_F^\top \mathbf{W} \mathbf{Q}_F \big]^{-1} \mathbf{Q}_F^\top \mathbf{W}\hat{\mathbf{q}}\,,
\end{align}
where $\mathbf{W} = \widehat{\bm{\Sigma}}^{-1}$ is the weighting matrix. By standard asymptotic arguments, the estimator $\hat{\sigma}$ (our primary object of interest) is asymptotically normal.

For stable distributions, the estimator facilitates inference on the tail index $\alpha$ via the transformation $\hat{\alpha} = \log 2 / \log \hat{\sigma}$. Although finite-sample variability may occasionally yield estimates outside the valid range $(0, 2]$, applying an optional range restriction ensures domain validity. This issue arises primarily when $\alpha$ is near $2$, i.e., when the tails are not particularly heavy, which is not a central concern in this paper.

Importantly, this section assumes access to two independent samples satisfying a location-scale (two-sum closure) model. In practice, however, only a single sample is available. The next section explores how to handle this case using a split-sample construction.

\section{Estimation via Sample Splitting}\label{sec: Sample Spliting}

\subsection{Two-Sample Construction via Random Assignment}\label{sec:Spliting}

We now consider the case in which only a single sample is available from a population that satisfies the two-sum closure property. While the preceding section assumed access to two independent samples, we extend the two-sample estimation framework to a one-sample setting by constructing two empirical distributions from a single observed dataset.

The key idea is to define two empirical distribution functions (EDFs), denoted $\hat{F}_1$ and $\hat{F}_2$, which serve as empirical analogs of the true \textit{baseline} and \textit{convolution} distributions, hereafter $F_1$ and $F_2$. The baseline distribution $F_1$ corresponds to the distribution of the original observations, while the convolution distribution $F_2$ corresponds to the distribution of pairwise sums of independent observations drawn from the same population. This construction enables plug-in estimation of parameters defined as functionals of $F_1$ and $F_2$. Our approach is motivated by \citet{kravitz2019sample}, who employed sample splitting in a different context involving stacked estimating equations and partitioning across distinct estimating functions.

Formally, suppose we observe $Z_1,\ldots,Z_n$ drawn \textit{i.i.d.} from a population satisfying the two-sum closure property. To construct baseline and convolution samples, we introduce $\delta_1, \ldots, \delta_n$ be \textit{i.i.d.} Bernoulli random variables with success probability $p \in (0,1)$. Each observation $Z_i$ is assigned to the baseline sample if $\delta_i = 1$ and to the convolution sample if $\delta_i = 0$. 

The resulting baseline sample $\mathcal{X}$ thus consists of observations on the original scale, while the convolution sample $\mathcal{Y}$ consists of all pairwise sums among observations assigned to the convolution group. Formally, we write the empirical samples as
\begin{align}   
    \mathcal{X} = \{Z_i : \delta_i = 1\} \quad \text{and} \quad \mathcal{Y} = \{Z_i + Z_j : \delta_i = \delta_j = 0, \; i < j \}.
\end{align}
These sets form the basis for constructing EDFs that approximate the distributional behavior of the baseline and convolution components. 

To this end, let 
\begin{align}
    n_1 = \sum_{i=1}^n \delta_i \quad \text{and}\quad n_2 = \sum_{1 \le i<j \le n} (1-\delta_i)(1-\delta_j)
\end{align}
denote the number of observations in the baseline and convolution groups. Because the assignments are random, $n_1$ and $n_2$ are random quantities. 

From the samples $\mathcal{X}$ and $\mathcal{Y}$, we next define empirical counting measures 
\begin{align}
    \hat{N}_1(x) = \sum_{i=1}^{n} \delta_i \mathbb{I}(Z_i \leq x)\quad \text{and}\quad \hat{N}_2(y) = \sum_{i<j} (1 - \delta_i)(1 - \delta_j) \mathbb{I}(Z_i + Z_j \leq y). \label{eq:counting_proc}
\end{align}
which count the number of observations in $\mathcal{X}$ and $\mathcal{Y}$ less than or equal to $x$ and $y$, respectively. 

The corresponding EDFs are
\begin{equation}
\hat{F}_1(x) = \frac{\hat{N}_1(x)}{n_1} \mathbb{I}(n_1 > 0)\quad \text{and}\quad \hat{F}_2(y) = \frac{\hat{N}_2(y)}{n_2} \mathbb{I}(n_2 > 0), \label{eq:split_edf_def}
\end{equation}
with EDFs set to zero if their respective denominators are zero. These functions serve as empirical approximations to $F_1$ and $F_2$.

Appendix \ref{sec:Appendix} establishes that $\hat{F}_1$ and $\hat{F}_2$ are consistent estimators of $F_1$ and $F_2$, respectively, with variances that decrease at the standard asymptotic rate $\text{O}\big(n^{-1}\big)$. These properties motivate the use of $\hat{F}_1$ and $\hat{F}_2$ as inputs in any two-sample estimation approach where the target parameter, viewed as a functional $\theta=\theta(F_1,F_2)$, is sufficiently smooth to admit asymptotically linear expansions. 

More generally, when the parameter of interest can be expressed as a functional $\theta=\theta(F_1,F_2)$, a natural estimator is the plug-in estimator $\hat{\theta}=\theta\big(\hat{F}_1,\hat{F}_2\big)$. \textcolor{black}{Under standard regularity conditions, a first-order functional expansion gives
\begin{align}
\theta\big(\hat{F}_1,\hat{F}_2\big)-\theta\big(F_1,F_2\big)
=
h_1\big(\hat{F}_1-F_1\big)
+
h_2\big(\hat{F}_2-F_2\big)
+
o_p\big(n^{-1/2}\big),
\end{align}
where $h_1$ and $h_2$ denote the corresponding first-order linear functionals. In other words, to first order, the estimation error in $\hat{\theta}$ is determined by the deviations of the empirical distribution functions $\hat{F}_1$ and $\hat{F}_2$ from their population counterparts.} This representation follows from the functional delta method (or, equivalently, a first-order von Mises expansion); see \citet[Chapter~21]{van2000} for a comprehensive treatment.

As an example, consider the WLS estimator of Section~\ref{sec:location-scale}. To apply it in the one-sample setting, we proceed by (a) randomly splitting the observations into baseline and convolution samples $\mathcal{X}$ and $\mathcal{Y}$ via independent Bernoulli trials, and then (b) applying the two-sample location-scale method to estimate $\sigma$ using the resulting empirical distributions. While this single-split approach is valid, it can be improved. The next section explores how repeated random splitting can enhance estimator stability and efficiency through aggregation.

\subsection{Repeated Splitting and Aggregation}

A single random partition may result in an estimator with elevated Monte Carlo variance, owing to substantial variability in the composition of the baseline and convolution samples induced by random assignment. To mitigate concern, we adopt a repeated-splitting strategy in which multiple independent random partitions are generated, and the corresponding estimators are averaged. This approach improves stability through aggregation while retaining the desirable properties of the individual plug-in estimators. We consider the implications of this strategy in this section.

Let $\delta_{bi}$, for $b = 1, \ldots, B$ and $i = 1, \ldots, n$, denote independent Bernoulli random variables with success probability $p$, where $B$ is the number of times sample splitting is applied. For each index $b$, the vector $[\delta_{b1}, \ldots, \delta_{bn}]^\top$ defines the random partition used in the $b^\text{th}$ split. Let $\mathcal{X}_b$ and $\mathcal{Y}_b$ denote the baseline and convolution samples corresponding to that split.

Suppose the target is a parameter vector $\bm{\theta} = [\theta_1, \ldots, \theta_K]^\top$ to be estimated via the two-sample methodology. In our application, this vector consists of the same parameter, $\sigma$ or $\alpha$, which is estimated using different choices of the spacing vectors $\mathbf{t}_k$ in the WLS framework. Let $\widehat{\bm{\theta}}_b = [\hat{\theta}_{b1}, \ldots, \hat{\theta}_{bK}]^\top$ denote the estimator obtained from the $b^\text{th}$ split, based on the corresponding samples $\mathcal{X}_b$ and $\mathcal{Y}_b$. 

Note that while the observations within each split—namely, those in $\mathcal{X}_b$ and $\mathcal{Y}_b$—are independent by construction, this independence does not hold across splits. In particular, the samples $\mathcal{X}_b$ and $\mathcal{X}_{b'}$ are not independent for $b \ne b'$, due to shared use of the underlying observations. For example, suppose for illustrative purposes that each $X_i$ has finite mean $\mu$ and variance $\kappa^2$. Then, for fixed $i$, the covariance between its contributions to two baseline samples is $\mathbb{C}\mathrm{ov}(\delta_{bi} X_i,\, \delta_{b'i} X_i) = p^2 \kappa^2$. This implies dependence between $\widehat{\bm{\theta}}_b$ and $\widehat{\bm{\theta}}_{b'}$ for $b\neq b'$. 

Define the aggregated estimator as the average over all $B$ splits
\begin{equation}
    \widehat{\bm{\theta}} = \frac{1}{B}\sum_{b=1}^{B} \widehat{\bm{\theta}}_b.
    \label{eq:average_est_over_splits}
\end{equation}
Then the variance of the aggregate estimator can be decomposed as
\begin{equation}
    \mathbb{V}\mathrm{ar}\big(\,\widehat{\bm{\theta}}\,\big) = \frac{1}{B}\,\bm{\Sigma} + \frac{B-1}{B}\,\bm{\Gamma},
    \label{eq:var_over_splits}
\end{equation}
where $\bm{\Sigma} = \mathbb{V}\mathrm{ar}\big(\widehat{\bm{\theta}}_b\big)$ and $\bm{\Gamma} = \mathbb{C}\mathrm{ov}\big(\widehat{\bm{\theta}}_b,\widehat{\bm{\theta}}_{b'}\big)$
with $b\neq b'$.

This decomposition highlights a key trade-off: as $B\rightarrow \infty$, the variance contribution from within-split variability, $\bm{\Sigma}$, vanishes, but the contribution from inter-split dependence, $\bm{\Gamma}$, persists and ultimately dominates. Accurate estimation of this covariance structure is therefore essential for valid inference. The following section develops a bootstrap-based procedure to estimate $\mathbb{V}\mathrm{ar}\big(\,\widehat{\bm{\theta}}\,\big)$ in the presence of repeated-splitting dependence.

\subsection{Bootstrap Variance Estimation via Extrapolation} \label{sec:bootstrap_extrap}

A natural approach for estimating the variance of the repeated-splitting estimator is to apply the bootstrap. However, a naïve implementation would require recomputing the estimator with $B$ sample splits for each bootstrap replicate, incurring substantial computational cost. Fortunately, Equation~\eqref{eq:var_over_splits} suggests a more efficient alternative. Recall that the variance of the aggregated estimator can be expressed as a linear combination of the within-split covariance $\bm{\Sigma}$ and the between-split covariance $\bm{\Gamma}$,
\begin{align}
    \mathbb{V}\mathrm{ar}\big(\,\widehat{\bm{\theta}}\,\big) = \frac{1}{B}\, \bm{\Sigma} + \frac{B-1}{B}\, \bm{\Gamma}.
\end{align}
Motivated by this decomposition, we estimate $\mathbb{V}\mathrm{ar}\big(\,\widehat{\bm{\theta}}\,\big)$ using a two-step procedure: first, estimate $\bm{\Sigma}$ and $\bm{\Gamma}$ based on a modest number of sample splits; second, use these estimatates to extrapolate the variance at the desired number of splits $B$.

Specifically, let $\widehat{\bm{\Omega}}_{B_1}$ and $\widehat{\bm{\Omega}}_{B_2}$ denote variance estimates computed from $B_1$ and $B_2$ sample splits, with $B_1 < B_2 < B$. These satisfy
\begin{align}\label{eq:Omega_over_LesSplits}
    \widehat{\bm{\Omega}}_{B_1} = \frac{1}{B_1} \widehat{\bm{\Sigma}} + \frac{B_1-1}{B_1} \widehat{\bm{\Gamma}} \quad \text{and}\quad
    \widehat{\bm{\Omega}}_{B_2} = \frac{1}{B_2} \widehat{\bm{\Sigma}} + \frac{B_2-1}{B_2} \widehat{\bm{\Gamma}}.
\end{align}

Treating $\bm{\Sigma}$ and $\bm{\Gamma}$ as unknowns, this forms a system of linear equations that can be solved explicitly. Solving to find an estimator of $\bm{\Gamma}$ yields
\begin{equation}
    \widehat{\bm{\Gamma}}_{(B_1,B_2)} = \frac{B_2\, \widehat{\bm{\Omega}}_{B_2} - B_1\, \widehat{\bm{\Omega}}_{B_1}}{B_2 - B_1}.
\end{equation}
Substituting into \eqref{eq:Omega_over_LesSplits} gives
\begin{equation}
    \widehat{\bm{\Sigma}}_{(B_1,B_2)} = B_1\, \widehat{\bm{\Omega}}_{B_1} - (B_1-1)\, \widehat{\bm{\Gamma}}_{(B_1,B_2)}.
\end{equation}
The subscripts emphasize dependence on the choice of integers $B_1$ and $B_2$. 

If the bootstrap is performed using only $B_{\text{max}}\ll B$ splits, there are still many valid $(B_1,B_2)$ pairs available. To fully leverage this information, we define overall estimators by averaging,
\begin{equation}
    \widehat{\bm{\Gamma}} = \frac{2}{(B_{\text{max}}-1)(B_{\text{max}}-2)}\sum_{b_1 = 2}^{B_{\text{max}}-1} \sum_{b_2 = b_1 + 1}^{B_{\text{max}}} \widehat{\bm{\Gamma}}_{(b_1,b_2)}, 
\end{equation}
with $\widehat{\bm{\Sigma}}$ obtained analogously by averaging over the possible $\widehat{\bm{\Sigma}}_{(b_1,b_2)}$.

Given these estimates, the variance at the targer number of $B$ splits is then extrapolated as
\begin{equation}
    \widetilde{\bm{\Omega}}_B = \frac{1}{B}\, \widehat{\bm{\Sigma}} + \frac{B-1}{B}\, \widehat{\bm{\Gamma}}.
\end{equation}

This procedure avoids the need for recomputing $B$ splits within each bootstrap replicate. Instead, it uses a smaller number of splits, $B_{\text{max}}$, and extrapolates the variance structure to the desired level $B$. Provided that $\widehat{\bm{\Omega}}_{B_1}$ and $\widehat{\bm{\Omega}}_{B_2}$ are consistent for all $B_1 < B_2 \leq B_{\text{max}}$, the resulting variance estimate is likewise consistent.

This strategy yields substantial computational savings when $B$ is large. Simulation results (Section~\ref{sec: Simulations}) illustrate this performance. In our experiments, $B_{\text{max}}=10$ provided reasonable accuracy in many cases, while $B_{\text{max}}=25$ consistently delivered strong performance. The ultimate choice should be guided by the computational resources available and, to a lesser extent, by the problem's sample size.

\section{Simulation Studies}\label{sec: Simulations}

\subsection{\textcolor{black}{Stability and Performance of the WLS Estimator}}

\textcolor{black}{The proposed tail index is naturally expressed as the scale parameter $\sigma$ in our location-scale convolution framework, whereas stable distributions generally make use of the $\alpha$ parameterization. For these distributions, $\sigma = 2^{1/\alpha}$, or equivalently, $\alpha = \log 2 / \log \sigma$. All results in this simulation study are reported on the $\sigma$ scale. This distinction is important when comparing reported results with other studies, for example, empirical work in \citet{mcculloch1986} and \citet{nolan2001}.}

Implementation of the split-sample WLS estimator presented in Section \ref{sec:location-scale} requires three choices from the user: the number of random splits $B$, the Bernoulli success probability $p$ for split allocation, and the number of quantiles $k$ in the WLS estimator, assuming throughout that these are equally spaced on the probability scale. This section presents a broad exploration of the impact of these choices on the estimator's accuracy.

\subsubsection{Choice of the Number of Splits}

The number of times sample splitting is applied can have a significant impact on the estimator's performance. If $B$ is too small, the resulting estimator exhibits high variability. On the other hand, a very large $B$ incurs considerable computational cost, detracting from the approach's practical utility. Therefore, selecting $B$ to balance these competing interests is essential. 

To assess the finite-sample impact of $B$ on estimation accuracy, we generated samples from standardized stable distributions (i.e., $\gamma=1$ and $\delta=0$ in \eqref{eq:cfk0}). Sample sizes $n \in \left\{250,500,1000\right\}$ were crossed with stability indices $\alpha \in \{0.4,0,5,\ldots,1.1,1.2\}$ and shape parameters $\beta \in \left\{0,0.25,0.75\right\}$. Random splits were generated using independent Bernoulli trials with $p=0.5$ as described in Section~\ref{sec:Spliting}. A total of $1000$ samples were generated under each of the $3\times 9 \times 3 = 81$ simulation conditions.

The WLS estimator was applied to each sample using equispaced $t_j=j/(k+1)$, $j=1,\ldots,k$ for number of quantiles $k\in\{5,6,...,16\}$. \textcolor{black}{The notation $\mathbf{t}_k$ is hereafter used to denote these quantile levels equally spaced on the probability scale.} Each estimator was evaluated for a range of split counts, $B \in \{
  1,\allowbreak
  2,\allowbreak
  3,\allowbreak
  4,\allowbreak
  5,\allowbreak
  10,\allowbreak
  20,\allowbreak
  30,\allowbreak
  40,\allowbreak
  50,\allowbreak
  100,\allowbreak
  150,\allowbreak
  200,\allowbreak
  250,\allowbreak
  500,\allowbreak
  1000\}$. 

\begin{figure}
    \centering
    \includegraphics[width=0.9\linewidth]{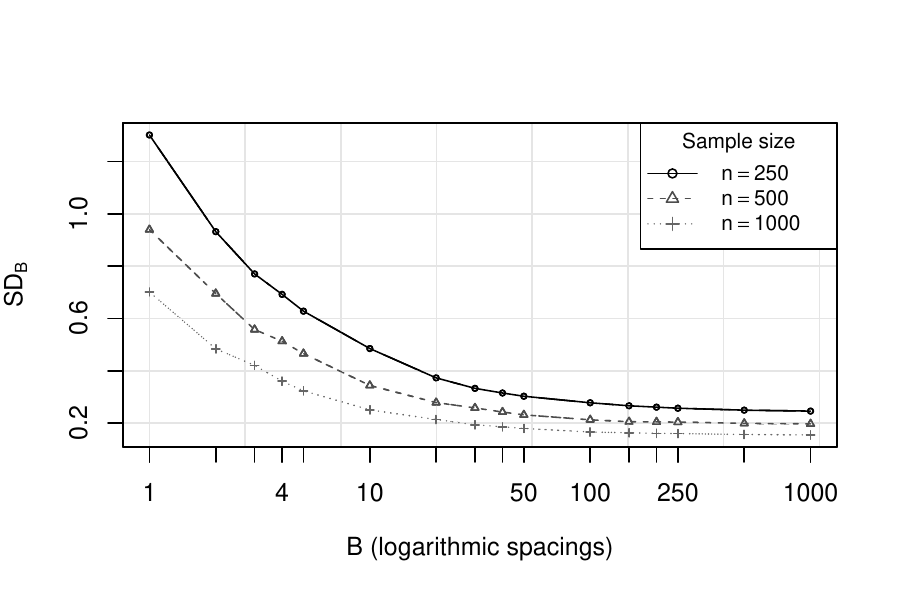}
    \caption{Impact of $B$ on WLS-$\mathbf{t}_{16}$ estimator for underlying $(\alpha,\beta)=(0.6,0)$ distribution.}
    \label{fig:effect_of_B}
\end{figure}

Figure~\ref{fig:effect_of_B} illustrates the impact of increasing $B$ for the case $(\alpha,\beta)=(0.6,0)$; similar behavior was observed across other parameter configurations. The standard deviation of the estimator, denoted $\mathrm{SD}_{\mathrm{B}}$, decreases steadily from $B=1$ to $B=100$, after which further gains become modest. By $B=250$, a plateau is evident, and further increases yield diminishing returns relative to additional computation time. For all remaining simulation studies in the paper, we adopt $B=250$, which offers a favorable balance between computational efficiency and estimator stability. Notably, achieving substantially lower standard deviations would require values of $B$ an order of magnitude larger. 

\FloatBarrier

\subsubsection{\textcolor{black}{Sensitivity to the Bernoulli Success Probability}}

Having established some empirical support for using $B=250$ splits, we next consider the role of the Bernoulli success probability $p$. Recall that this parameter determines how likely an observation is to be assigned to the sample retained on the original scale, yielding approximately $np$ observations, rather than the sample undergoing empirical convolution, containing approximately $n(1-p)\times\big\{n(1-p)-1\big\}/2$ observations after taking pairwise sums. A larger $p$ keeps more observations on the original scale per split but results in fewer on the convolution scale. Note also that these pairwise sums are dependent, so a larger number of them may not yield a proportional increase in the information available for estimation.

For this investigation, we report results for sample size $n=500$ using the WLS-$\mathbf{t}_10$ estimator, noting that sample sizes $n\in\{250,1000\}$ and number of quantiles $k\in\{5,16\}$ gave qualitatively similar results. We evaluated the estimators using allocation success probabilities $p\in\{0.20,0.35,0.50,0.65,0.80\}$. Samples were generated from standardized stable distributions with $\alpha\in\{0.4,\ldots,1.9\}$ and $\beta\in\{0,0.75,1\}$. A total of $R=1000$ samples were generated for each of the $16\times 3 = 48$ $(\alpha,\beta)$ configurations. For values $(\alpha,\beta,p)$, we define empirical root mean squared error (RMSE) as
$$\mathrm{RMSE}_{\alpha,\beta}(p) = \sqrt{\frac{1}{R}\sum_{r=1}^{R} (\hat\sigma_r - \sigma)^2}.$$

Table \ref{tab: RMSE_p_sensitivity} illustrates two specific cases, chosen to demonstrate that there is no uniform optimal choice of $p$. For $(\alpha,\beta) = (0.4,0)$, the smallest RMSE is obtained when $p=0.35$, while for $(\alpha,\beta) = (1.9,1)$, the minimum occurs at $p=0.65$ (and noting that the RMSE at $p=0.8$ is practically indistinguishable given the Monte Carlo variability).

\begin{table}
\centering\small
\begin{tabular}{cccccc}
\toprule
\multicolumn{1}{c}{$(\alpha,\beta)$} &
\multicolumn{1}{c}{$p=0.20$} &
\multicolumn{1}{c}{$p=0.35$} &
\multicolumn{1}{c}{$p=0.50$} &
\multicolumn{1}{c}{$p=0.65$} &
\multicolumn{1}{c}{$p=0.80$} \\
\midrule
$(0.4,\ 0)$   & 0.755 & 0.692 & 0.708 & 0.699 & 0.896 \\
$(1.9,\ 1)$   & 0.039 & 0.029 & 0.023 & 0.022 & 0.022 \\
\bottomrule
\end{tabular}
\caption{RMSE of the WLS-$\mathbf{t}_{10}$ estimator of $\sigma$ as a function of the split fraction $p$, for two $(\alpha,\beta)$ configurations.}
\label{tab: RMSE_p_sensitivity}
\end{table}

To evaluate performance over all $(\alpha,\beta)$ configurations, we define the \emph{relative RMSE}, using $p = 0.5$ as the baseline, as
$$
\mathrm{rRMSE}_{\alpha,\beta}(p) = \frac{\mathrm{RMSE}_{\alpha,\beta}(p)}{\mathrm{RMSE}_{\alpha,\beta}(0.5)}.
$$
The averaged relative RMSE curve is found by averaging over the $48$ $(\alpha,\beta)$ configurations considered,
$$
\overline{\mathrm{rRMSE}}(p) = \frac{1}{48} \sum_{(\alpha,\beta)} \mathrm{rRMSE}_{\alpha,\beta}(p).
$$

\begin{figure}[htbp]
\centering
\includegraphics[width=0.9\textwidth]{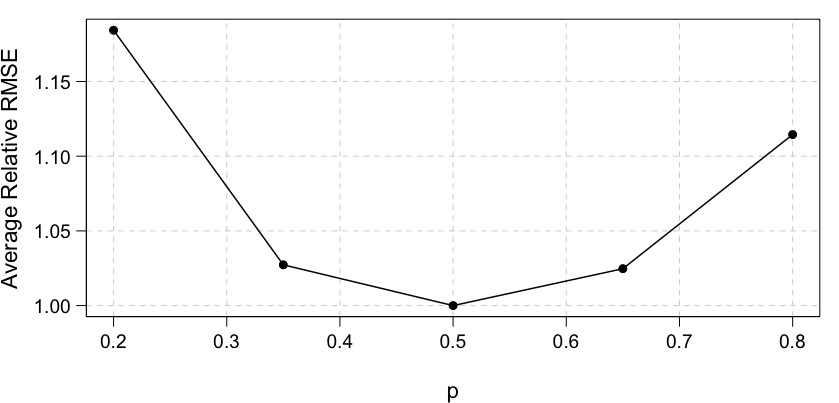}
\caption{Average relative RMSE as a function of the split fraction $p$, averaged across all $(\alpha,\beta)$ configurations, with $p = 0.5$ as the baseline.}
\label{fig:rel_rmse_p}
\end{figure}

While this summary masks variation across individual parameter configurations, we are encouraged to find that $p=0.5$ minimizes the average relative RMSE. The more extreme choices, here $p\in\{0.2,0.8\}$, yield the largest average relative RMSE. These results provide some empirical support for using $p=0.5$, especially since any improvement from a different choice appears to depend on the unknown model parameters. It may be possible to select $p$ adaptively, but this investigation falls outside the scope of our present study.

\subsubsection{Choosing the Number of Quantiles}

\begin{figure}
    \centering
    \includegraphics[width=0.9\linewidth]{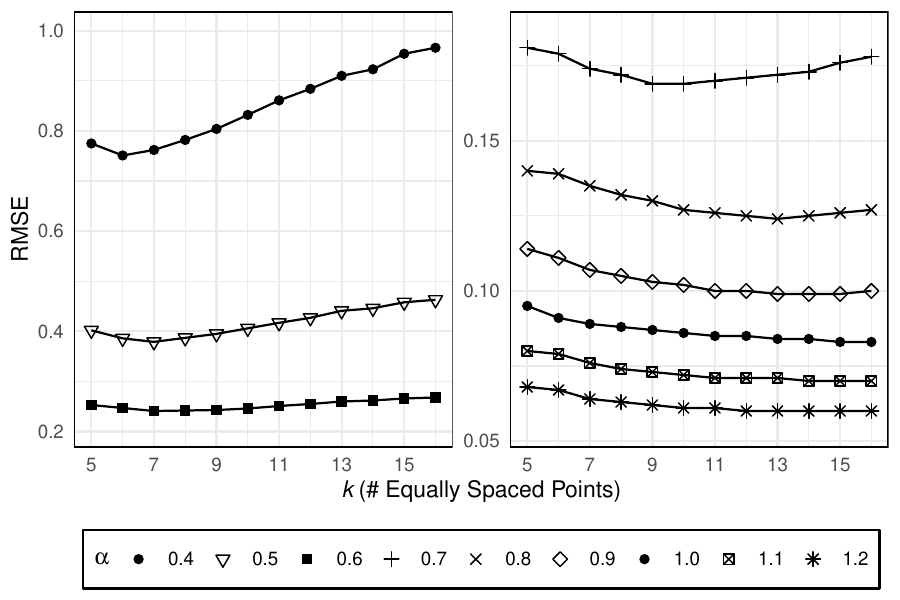}
    \caption{RMSE of the WLS estimators for $\mathbf{t}_k$, $k=5,\ldots,16$, with $\beta=0$ and $n=500$.}
    \label{fig:WLS_RMSE_symmetric}
\end{figure}

\begin{figure}
    \centering
    \includegraphics[width=0.9\linewidth]{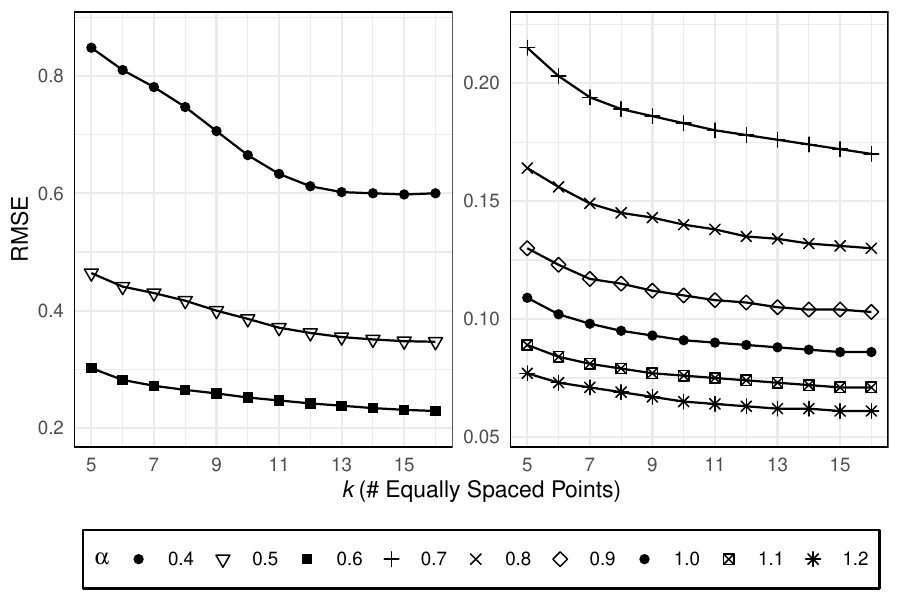}
    \caption{RMSE of the WLS estimators for $\mathbf{t}_k$, $k=5,\ldots,16$, with $\beta=0.75$ and $n=500$.}
    \label{fig:WLS_RMSE_skew}
\end{figure}

Next, we explore the impact of the number of quantiles used in the WLS estimator. Figures \ref{fig:WLS_RMSE_symmetric} and \ref{fig:WLS_RMSE_skew} present the RMSE of these estimators with the equispaced $\mathbf{t}_k$ for $k\in\{5,\ldots,16\}$ and $n=500$ under symmetric ($\beta=0$) and skewed ($\beta=0.75$) scenarios, respectively. The results for $n = 250$ and $n = 1000$ are qualitatively similar and omitted for brevity.

Figure \ref{fig:WLS_RMSE_symmetric} shows that in the symmetric case, smaller values of $k$ can outperform larger ones when $\alpha = 0.4$, corresponding to extremely heavy-tailed distributions. For $\alpha \in \{0.5, \ldots, 0.9\}$, smaller $k$ values still perform competitively, although the differences are less pronounced. When $\alpha \geq 1$, performance improves with $k$, with $k = 16$ yielding the best results. 

In contrast, Figure~\ref{fig:WLS_RMSE_skew} shows that under skewness, all estimators benefit from increasing $k$. Even for $\alpha = 0.4$, RMSE decreases noticeably with larger $k$, indicating that a ``greedy'' use of quantiles is beneficial in asymmetric settings.

Based on these findings, we consider $k \in \{5, 10, 16\}$ as representative choices in the broader method-comparison simulations that follow.

\subsubsection{Comparison with Existing Estimators}

The split-sample WLS approach \textcolor{black}{with equispaced $\mathbf{t}_k$} is compared to several widely used tail index estimators for stable distributions: the characteristic function method of \citet{koutr1980} (hereafter, the Koutrouvelis method); the sample quantile matching method of \citet{mcculloch1986} (hereafter, the McCulloch estimator); and the maximum likelihood estimator (MLE), all implemented via the \texttt{StableEstim} package in R. 

\textcolor{black}{As the WLS approach directly estimates the scale parameter $\sigma$ arising from the location-scale representation, we use this scale as the primary basis for comparison. The comparator methods estimate $\alpha = \log 2 / \log \sigma$, and we therefore transform the latter using $\sigma = 2^{1/\alpha}$. This places all estimators on the scale directly targeted by our procedure. Summary results on the $\alpha$ scale are provided in the GitHub repository accompanying this paper.}

\textcolor{black}{Before presenting the simulation results, it is useful to illustrate why comparisons on the $\alpha$ and $\sigma$ scales differ substantively. Let $$g(\sigma) = \frac{\log 2}{\log \sigma}$$ denote the transformation to the scale of interest here, giving a first-order Taylor expansion
$$\hat{\alpha} - \alpha = g'(\sigma)\, (\hat{\sigma} - \sigma) +
o_p\big(|\hat{\sigma}-\sigma|\big),$$
where $$g'(\sigma) = -\frac{\alpha^2}{\sigma\log 2}.$$
Consequently, for interior parameter values and large enough sample sizes, it follows that 
$$\operatorname{RMSE}(\hat{\alpha})
\approx
\frac{\alpha^2}{\sigma\log 2}
\operatorname{RMSE}(\hat{\sigma}).$$
A consequence of this is that a small RMSE value on the $\alpha$ scale may correspond to a much larger RMSE on the $\sigma$ scale, especially when $\alpha$ is small.}

\textcolor{black}{The method-comparison simulation was done under a broad range of stable distributions. Specifically, we used tail indices $\alpha\in\{0.20,\allowbreak 0.40,\allowbreak 0.60,\allowbreak 0.80,\allowbreak 1.00,\allowbreak 1.20,\allowbreak 1.40,\allowbreak 1.60,\allowbreak 1.80,\allowbreak 1.95\}$ corresponding to, after rounding, to $\sigma \in \{32.00,\allowbreak 5.66,\allowbreak 3.17,\allowbreak 2.38,\allowbreak 2.00,\allowbreak 1.78,\allowbreak 1.64,\allowbreak 1.54,\allowbreak 1.47,\allowbreak 1.43\}$, together with skewness parameters $\beta\in\{0,0.75,1\}$ and sample sizes $n\in\{250,500,1000\}$. The value $\alpha = 1.95$ is included to study the behavior near the Gaussian limit without actually considering $\alpha = 2$, where the skewness parameter no longer has a meaningful interpretation. For each of the $10\times 3 \times 3 = 90$ simulation configurations, $1000$ samples were generated from the appropriate stable distributions. Estimators were computed for all samples, except for the MLE due to its substantially higher computational burden (minutes per run vs. seconds). Thus, MLE was evaluated on $250$ of the $1000$ samples in each instance.}

Tables~\ref{tab: RMSE_n250}, \ref{tab: RMSE_n500}, and \ref{tab: RMSE_n1000} report the $\sigma$-scale RMSE values for each estimator after trimming the largest $5\%$ of squared errors. This choice was motivated primarily by the occasional extreme outliers produced by the MLE, and was then applied uniformly to all estimators for consistency and comparability. The corresponding untrimmed RMSE values on the $\sigma$ scale, as well as trimmed and untrimmed $\alpha$ scale results, are all available on the accompanying GitHub repository. 

As a reminder, the MCCulloch estimator was developed for $\alpha\geq 0.5$, see \citet{mcculloch1986}. In our numerical experiments, the existing implementation of this method in the \texttt{R} library \texttt{StableEstim} failed to give a result in over $99\%$ of the samples generated for $\alpha \leq 0.4$. Also interestingly, it failed in around $40\%$ of cases with $\alpha = 1.95$. We consequently made the decision to only report performance metrics for this estimator when more than $90\%$ of the generated datasets resulted in a converged parameter estimate.

\begin{table}
\centering\small
\begin{tabular}{crrrrrr}
\toprule
\multicolumn{1}{c}{} & \multicolumn{6}{c}{$\beta = 0$} \\
\cmidrule{2-7}
\multicolumn{1}{c}{$\alpha$} & 
\multicolumn{1}{c}{$\mathbf{t}_5$} & 
\multicolumn{1}{c}{$\mathbf{t}_{10}$} & 
\multicolumn{1}{c}{$\mathbf{t}_{16}$} & 
\multicolumn{1}{c}{Kout} & 
\multicolumn{1}{c}{McC} & 
\multicolumn{1}{c}{MLE} \\
\hline
0.20 & 10.620 & 12.117 & 13.720 & 47.105 & -- & 20.579 \\
0.40 & 0.902 & 0.996 & 1.099 & 1.228 & -- & 0.591 \\
0.60 & 0.301 & 0.298 & 0.309 & 0.400 & -- & 0.215 \\
0.80 & 0.166 & 0.152 & 0.150 & 0.158 & 0.205 & 0.120 \\
1.00 & 0.116 & 0.108 & 0.101 & 0.102 & 0.134 & 0.085 \\
1.20 & 0.083 & 0.074 & 0.071 & 0.071 & 0.086 & 0.068 \\
1.40 & 0.065 & 0.055 & 0.053 & 0.056 & 0.066 & 0.049 \\
1.60 & 0.050 & 0.045 & 0.042 & 0.038 & 0.053 & 0.042 \\
1.80 & 0.042 & 0.034 & 0.032 & 0.024 & -- & 0.026 \\
1.95 & 0.037 & 0.027 & 0.025 & 0.013 & -- & 0.010 \\
\toprule
\multicolumn{1}{c}{} & \multicolumn{6}{c}{$\beta = 0.75$} \\
\cmidrule{2-7}
\multicolumn{1}{c}{$\alpha$} & 
\multicolumn{1}{c}{$\mathbf{t}_5$} & 
\multicolumn{1}{c}{$\mathbf{t}_{10}$} & 
\multicolumn{1}{c}{$\mathbf{t}_{16}$} & 
\multicolumn{1}{c}{Kout} & 
\multicolumn{1}{c}{McC} & 
\multicolumn{1}{c}{MLE} \\
\hline
0.20 & 22.185 & 8.557 & 7.805 & 52.320 & -- & 17.207 \\
0.40 & 1.145 & 0.843 & 0.737 & 0.857 & -- & 0.482 \\
0.60 & 0.405 & 0.333 & 0.306 & 0.325 & 0.360 & 0.198 \\
0.80 & 0.208 & 0.185 & 0.167 & 0.147 & 0.229 & 0.119 \\
1.00 & 0.134 & 0.118 & 0.113 & 0.095 & 0.142 & 0.080 \\
1.20 & 0.093 & 0.081 & 0.077 & 0.068 & 0.098 & 0.057 \\
1.40 & 0.072 & 0.063 & 0.060 & 0.058 & 0.069 & 0.043 \\
1.60 & 0.054 & 0.045 & 0.043 & 0.037 & 0.048 & 0.031 \\
1.80 & 0.046 & 0.037 & 0.034 & 0.025 & -- & 0.025 \\
1.95 & 0.038 & 0.029 & 0.026 & 0.014 & -- & 0.011 \\
\toprule
\multicolumn{1}{c}{} & \multicolumn{6}{c}{$\beta = 1$} \\
\cmidrule{2-7}
\multicolumn{1}{c}{$\alpha$} & 
\multicolumn{1}{c}{$\mathbf{t}_5$} & 
\multicolumn{1}{c}{$\mathbf{t}_{10}$} & 
\multicolumn{1}{c}{$\mathbf{t}_{16}$} & 
\multicolumn{1}{c}{Kout} & 
\multicolumn{1}{c}{McC} & 
\multicolumn{1}{c}{MLE} \\
\hline
0.20 & 19.853 & 14.324 & 12.374 & 57.545 & -- & 43.584 \\
0.40 & 1.141 & 0.914 & 0.850 & 1.085 & -- & 0.451 \\
0.60 & 0.395 & 0.336 & 0.310 & 0.360 & 0.280 & 0.174 \\
0.80 & 0.212 & 0.183 & 0.174 & 0.146 & 0.191 & 0.091 \\
1.00 & 0.135 & 0.120 & 0.113 & 0.098 & 0.129 & 0.062 \\
1.20 & 0.098 & 0.086 & 0.082 & 0.076 & 0.093 & 0.056 \\
1.40 & 0.074 & 0.065 & 0.063 & 0.058 & 0.068 & 0.038 \\
1.60 & 0.058 & 0.049 & 0.046 & 0.038 & 0.050 & 0.031 \\
1.80 & 0.044 & 0.036 & 0.033 & 0.025 & -- & 0.019 \\
1.95 & 0.041 & 0.030 & 0.027 & 0.015 & -- & 0.013 \\
\bottomrule
\end{tabular}
\caption{5\% trimmed RMSE of various estimators of $\sigma = 2^{1/\alpha}$ for $n = 250$.}
\label{tab: RMSE_n250}
\end{table} 

\begin{table}
\centering\small
\begin{tabular}{crrrrrr}
\toprule
\multicolumn{1}{c}{} & \multicolumn{6}{c}{$\beta = 0$} \\
\cmidrule{2-7}
\multicolumn{1}{c}{$\alpha$} & 
\multicolumn{1}{c}{$\mathbf{t}_5$} & 
\multicolumn{1}{c}{$\mathbf{t}_{10}$} & 
\multicolumn{1}{c}{$\mathbf{t}_{16}$} & 
\multicolumn{1}{c}{Kout} & 
\multicolumn{1}{c}{McC} & 
\multicolumn{1}{c}{MLE} \\
\hline
0.20 & 9.730 & 10.504 & 12.000 & 24.034 & -- & 18.494 \\
0.40 & 0.693 & 0.701 & 0.823 & 0.850 & -- & 0.347 \\
0.60 & 0.211 & 0.211 & 0.225 & 0.232 & 0.218 & 0.137 \\
0.80 & 0.121 & 0.112 & 0.110 & 0.108 & 0.145 & 0.076 \\
1.00 & 0.079 & 0.076 & 0.074 & 0.070 & 0.095 & 0.058 \\
1.20 & 0.055 & 0.052 & 0.051 & 0.050 & 0.057 & 0.047 \\
1.40 & 0.044 & 0.039 & 0.039 & 0.039 & 0.042 & 0.034 \\
1.60 & 0.034 & 0.029 & 0.027 & 0.025 & 0.036 & 0.025 \\
1.80 & 0.028 & 0.022 & 0.020 & 0.018 & -- & 0.018 \\
1.95 & 0.026 & 0.018 & 0.016 & 0.010 & -- & 0.009 \\
\toprule
\multicolumn{1}{c}{} & \multicolumn{6}{c}{$\beta = 0.75$} \\
\cmidrule{2-7}
\multicolumn{1}{c}{$\alpha$} & 
\multicolumn{1}{c}{$\mathbf{t}_5$} & 
\multicolumn{1}{c}{$\mathbf{t}_{10}$} & 
\multicolumn{1}{c}{$\mathbf{t}_{16}$} & 
\multicolumn{1}{c}{Kout} & 
\multicolumn{1}{c}{McC} & 
\multicolumn{1}{c}{MLE} \\
\hline
0.20 & 10.770 & 5.981 & 5.914 & 25.319 & -- & 15.680 \\
0.40 & 0.701 & 0.573 & 0.529 & 0.519 & -- & 0.328 \\
0.60 & 0.245 & 0.220 & 0.202 & 0.199 & 0.274 & 0.138 \\
0.80 & 0.132 & 0.119 & 0.116 & 0.103 & 0.162 & 0.082 \\
1.00 & 0.088 & 0.075 & 0.072 & 0.066 & 0.102 & 0.059 \\
1.20 & 0.064 & 0.058 & 0.056 & 0.048 & 0.072 & 0.042 \\
1.40 & 0.050 & 0.042 & 0.040 & 0.038 & 0.051 & 0.033 \\
1.60 & 0.037 & 0.030 & 0.028 & 0.026 & 0.035 & 0.025 \\
1.80 & 0.031 & 0.023 & 0.021 & 0.017 & -- & 0.015 \\
1.95 & 0.026 & 0.019 & 0.016 & 0.010 & -- & 0.009 \\
\toprule
\multicolumn{1}{c}{} & \multicolumn{6}{c}{$\beta = 1$} \\
\cmidrule{2-7}
\multicolumn{1}{c}{$\alpha$} & 
\multicolumn{1}{c}{$\mathbf{t}_5$} & 
\multicolumn{1}{c}{$\mathbf{t}_{10}$} & 
\multicolumn{1}{c}{$\mathbf{t}_{16}$} & 
\multicolumn{1}{c}{Kout} & 
\multicolumn{1}{c}{McC} & 
\multicolumn{1}{c}{MLE} \\
\hline
0.20 & 9.989 & 7.517 & 6.225 & 23.218 & -- & 38.159 \\
0.40 & 0.687 & 0.589 & 0.533 & 0.708 & -- & 0.301 \\
0.60 & 0.258 & 0.222 & 0.198 & 0.218 & 0.224 & 0.119 \\
0.80 & 0.142 & 0.119 & 0.110 & 0.104 & 0.132 & 0.072 \\
1.00 & 0.095 & 0.082 & 0.078 & 0.070 & 0.088 & 0.050 \\
1.20 & 0.068 & 0.059 & 0.057 & 0.051 & 0.060 & 0.040 \\
1.40 & 0.051 & 0.045 & 0.042 & 0.039 & 0.047 & 0.030 \\
1.60 & 0.039 & 0.033 & 0.030 & 0.026 & 0.034 & 0.019 \\
1.80 & 0.030 & 0.023 & 0.021 & 0.017 & -- & 0.013 \\
1.95 & 0.027 & 0.018 & 0.015 & 0.011 & -- & 0.008 \\
\bottomrule
\end{tabular}
\caption{5\% trimmed RMSE of various estimators of $\sigma = 2^{1/\alpha}$ for $n = 500$.}
\label{tab: RMSE_n500}
\end{table} 

\begin{table}
\centering\small
\begin{tabular}{crrrrrr}
\toprule
\multicolumn{1}{c}{} & \multicolumn{6}{c}{$\beta = 0$} \\
\cmidrule{2-7}
\multicolumn{1}{c}{$\alpha$} & 
\multicolumn{1}{c}{$\mathbf{t}_5$} & 
\multicolumn{1}{c}{$\mathbf{t}_{10}$} & 
\multicolumn{1}{c}{$\mathbf{t}_{16}$} & 
\multicolumn{1}{c}{Kout} & 
\multicolumn{1}{c}{McC} & 
\multicolumn{1}{c}{MLE} \\
\hline
0.20 & 8.086 & 8.112 & 9.416 & 14.109 & -- & 13.013 \\
0.40 & 0.518 & 0.516 & 0.591 & 0.544 & -- & 0.229 \\
0.60 & 0.169 & 0.157 & 0.164 & 0.166 & -- & 0.105 \\
0.80 & 0.087 & 0.082 & 0.084 & 0.077 & 0.109 & 0.063 \\
1.00 & 0.057 & 0.052 & 0.053 & 0.048 & 0.072 & 0.040 \\
1.20 & 0.041 & 0.040 & 0.038 & 0.036 & 0.044 & 0.031 \\
1.40 & 0.030 & 0.027 & 0.026 & 0.025 & 0.031 & 0.023 \\
1.60 & 0.026 & 0.021 & 0.020 & 0.019 & 0.027 & 0.018 \\
1.80 & 0.021 & 0.016 & 0.015 & 0.014 & 0.024 & 0.014 \\
1.95 & 0.019 & 0.013 & 0.011 & 0.008 & -- & 0.006 \\
\toprule
\multicolumn{1}{c}{} & \multicolumn{6}{c}{$\beta = 0.75$} \\
\cmidrule{2-7}
\multicolumn{1}{c}{$\alpha$} & 
\multicolumn{1}{c}{$\mathbf{t}_5$} & 
\multicolumn{1}{c}{$\mathbf{t}_{10}$} & 
\multicolumn{1}{c}{$\mathbf{t}_{16}$} & 
\multicolumn{1}{c}{Kout} & 
\multicolumn{1}{c}{McC} & 
\multicolumn{1}{c}{MLE} \\
\hline
0.20 & 5.982 & 4.121 & 4.347 & 13.288 & -- & 12.962 \\
0.40 & 0.446 & 0.407 & 0.395 & 0.366 & -- & 0.227 \\
0.60 & 0.166 & 0.155 & 0.151 & 0.142 & 0.210 & 0.094 \\
0.80 & 0.094 & 0.086 & 0.084 & 0.072 & 0.114 & 0.060 \\
1.00 & 0.063 & 0.055 & 0.054 & 0.049 & 0.075 & 0.038 \\
1.20 & 0.046 & 0.040 & 0.038 & 0.035 & 0.054 & 0.031 \\
1.40 & 0.034 & 0.029 & 0.028 & 0.025 & 0.038 & 0.023 \\
1.60 & 0.026 & 0.022 & 0.021 & 0.020 & 0.028 & 0.019 \\
1.80 & 0.022 & 0.017 & 0.015 & 0.013 & 0.021 & 0.011 \\
1.95 & 0.019 & 0.013 & 0.011 & 0.008 & -- & 0.006 \\
\toprule
\multicolumn{1}{c}{} & \multicolumn{6}{c}{$\beta = 1$} \\
\cmidrule{2-7}
\multicolumn{1}{c}{$\alpha$} & 
\multicolumn{1}{c}{$\mathbf{t}_5$} & 
\multicolumn{1}{c}{$\mathbf{t}_{10}$} & 
\multicolumn{1}{c}{$\mathbf{t}_{16}$} & 
\multicolumn{1}{c}{Kout} & 
\multicolumn{1}{c}{McC} & 
\multicolumn{1}{c}{MLE} \\
\hline
0.20 & 6.341 & 4.832 & 4.227 & 14.636 & -- & 34.079 \\
0.40 & 0.441 & 0.398 & 0.371 & 0.448 & -- & 0.216 \\
0.60 & 0.174 & 0.152 & 0.146 & 0.152 & 0.165 & 0.090 \\
0.80 & 0.096 & 0.086 & 0.081 & 0.073 & 0.094 & 0.050 \\
1.00 & 0.066 & 0.056 & 0.053 & 0.048 & 0.060 & 0.034 \\
1.20 & 0.049 & 0.041 & 0.040 & 0.035 & 0.045 & 0.028 \\
1.40 & 0.038 & 0.032 & 0.029 & 0.027 & 0.035 & 0.021 \\
1.60 & 0.028 & 0.023 & 0.022 & 0.019 & 0.025 & 0.014 \\
1.80 & 0.022 & 0.016 & 0.015 & 0.013 & 0.021 & 0.011 \\
1.95 & 0.019 & 0.012 & 0.010 & 0.008 & -- & 0.006 \\
\bottomrule
\end{tabular}
\caption{5\% trimmed RMSE of various estimators of $\sigma = 2^{1/\alpha}$ for $n = 1000$.}
\label{tab: RMSE_n1000}
\end{table} 

When examining the results, MLE generally has the lowest RMSE, as one would expect from the asymptotically optimal approach. A fairly notable exception occurs at $\alpha=0.2$, where all three WLS estimators have lower RMSE than the MLE. It is unclear whether MLE failure reflects a finite-sample property or numeric instability with the existing implementation in \texttt{R}. This should be explored in future research. Regardless of the reason, the WLS approach gives an alternative approach when the tails are extremely heavy. For $\alpha\in\{0.4,0.6\}$, WLS also outperformed the Koutrouvelis and McCulloch estimators (when the latter results in valid estimates). While the WLS estimators' RMSE is larger than that of MLE, it does have a much lower computation time and does not require numeric inversion of the characteristic function.

The advantage of WLS over the Koutrouvelis approach and MLE diminishes as $\alpha$ increases, especially under skewed distributions. Section~\ref{sec:WLSoptimal} revisits the question of optimality and demonstrates how WLS can be further refined. These results should therefore be viewed as provisional rather than definitive.

\FloatBarrier

\subsection{Variance Estimation} \label{sec:var_est}

To evaluate the effectiveness of the reduced-split bootstrap procedure for variance estimation, we conducted a simulation study across 12 configurations defined by $(n, \alpha, \beta) \in \{250, 500, 1000\} \times \{0.6, 1.2\} \times \{0, 0.75\}$. For each configuration, $M = 500$ datasets of size $n$ were drawn from a standardized stable distribution with tail index $\alpha$ and asymmetry index $\beta$. For each dataset, three WLS estimators were computed using equispaced quantile vectors $\mathbf{t}_k$, $k \in \{5, 10, 16\}$. Bootstrap variance estimates were obtained within each simulation replicate using $n{\text{boot}} = 100$ resamples.

For arbitrary number of splits $B$ and each bootstrap replicate $v = 1, \ldots, n_{\text{boot}}$, let $\widehat{\bm{\theta}}_{v,B}^\ast$ denote a (possibly vector-valued) estimator computed using $B$ independent sample splits. Define the bootstrap mean as $\bar{\bm{\theta}}_B^\ast$ and the bootstrap covariance matrix as
\begin{align}
    \widehat{\bm{\Omega}}_{B} = \frac{1}{n_{\text{boot}}-1} \sum_{v=1}^{n_{\text{boot}}} \Big(\widehat{\bm{\theta}}^\ast_{v,B} - \bar{\bm{\theta}}_B^\ast\Big)\Big(\widehat{\bm{\theta}}^\ast_{v,B} - \bar{\bm{\theta}}_B^\ast\Big)^\top, 
\end{align}
where the subscript emphasizes dependence on the number of splits used.

We compared two variance estimation procedures. The full bootstrap replicated the entire set of $B = 250$ sample splits in each of the $n_{\text{boot}} = 100$ bootstrap iterations, yielding the empirical covariance estimate $\widehat{\bm{\Omega}}_{250}$. In contrast, the reduced-split bootstrap used only $B = 25$ splits per replicate and extrapolated to $B = 250$ using the variance decomposition framework from Section~\ref{sec:bootstrap_extrap}. Because performing $B$ splits inherently yields results for all smaller split counts $b < B$, each replicate also produced $\widehat{\bm{\Omega}}_b$ for all $2 \le b \le 25$ at minimal additional computational cost.

In particular, the reduced-split bootstrap leverages the fact that each replicate provides estimates $\widehat{\bm{\Omega}}_b$ for multiple values of $b$. Thus, within each bootstrap replicate, and for every pair $(b_1, b_2)$ with $2 \le b_1 < b_2 \le 25$, we estimate the between-split and within-split covariance matrices
\begin{align}
    \widehat{\bm{\Gamma}}_{(b_1,b_2)} = \frac{b_2 \widehat{\bm{\Omega}}_{b_2} - b_1 \widehat{\bm{\Omega}}_{b_1}}{b_2 - b_1}\quad \text{and}\quad
    \widehat{\bm{\Sigma}}_{(b_1,b_2)} = b_1 \widehat{\bm{\Omega}}_{b_1} - (b_1 - 1)\widehat{\bm{\Gamma}}_{(b_1,b_2)}. 
\end{align}
These calculations are repeated across all $n_{\text{boot}}$ replicates, and the resulting estimates are then averaged over the $276$ unique $(b_1, b_2)$ pairs to obtain final component estimates
\begin{align}
    \widehat{\bm{\Gamma}} = \frac{1}{276} \sum_{b_1 = 2}^{24} \sum_{b_2 = b_1 + 1}^{25} \widehat{\bm{\Gamma}}_{(b_1,b_2)} \quad \text{and}\quad
    \widehat{\bm{\Sigma}} = \frac{1}{276} \sum_{b_1 = 2}^{24} \sum_{b_2 = b_1 + 1}^{25} \widehat{\bm{\Sigma}}_{(b_1,b_2)}.
\end{align}

The extrapolated covariance estimator for $B = 250$ splits is then given by
\begin{align}
    \widetilde{\bm{\Omega}}_{250} = \frac{1}{250}\, \widehat{\bm{\Sigma}} + \frac{249}{250}\, \widehat{\bm{\Gamma}}.
\end{align}
This mimics the full bootstrap estimate $\widehat{\bm{\Omega}}_{250}$ at significantly reduced computational cost. For brevity, we suppress explicit reference to the number of splits used to compute $\widehat{\bm{\Sigma}}$ and $\widehat{\bm{\Gamma}}$.

For each of the $m=1,\ldots, M = 500$ datasets, both covariance estimates, $\widehat{\bm{\Omega}}_{250,m}$ and $\widetilde{\bm{\Omega}}_{250,m}$, were computed to facilitate comparison. Letting $\widehat{\bm{\Omega}}^{\mathrm{boot}}_{250,m}$ denote either method, we averaged across datasets to estimate the expected bootstrap variance under each approach.

Let $j = 1, 2, 3$ index the WLS estimators corresponding to $\mathbf{t}_5$, $\mathbf{t}_{10}$, and $\mathbf{t}_{16}$, respectively. Define
\begin{align}
    \widehat{\mathrm{Var}}^{\mathrm{boot}}_j = \frac{1}{M} \sum_{m=1}^M \left[\widehat{\bm{\Omega}}^{\mathrm{boot}}_{250,m}\right]_{j,j} \quad \text{and}\quad
    \widehat{\mathrm{Var}}^{\mathrm{MC}}_j = \frac{1}{M-1} \sum_{m=1}^M \left(\hat{\theta}_j^{(m)} - \bar{\theta}_j\right)^2,
\end{align}
where $\hat{\theta}_j^{(m)}$ is the $j$th component of the WLS estimator in dataset $m$, and $\bar{\theta}_j$ is its average across the $M$ replications. The \textit{variance ratio} is
\begin{align}
    \text{Ratio}_j = {\widehat{\mathrm{Var}}^{\mathrm{boot}}_j}/\, {\widehat{\mathrm{Var}}^{\mathrm{MC}}_j}, \quad j = 1, 2, 3.
\end{align}
This ratio assesses how well the bootstrap variance approximates the true sampling variance. Values near 1 suggest accurate estimation; values above or below 1 indicate over- or underestimation, respectively. Jackknife standard errors across the $M = 500$ datasets were computed to assess variability in these ratios. Table~\ref{tab:var_ratios} reports the results, with each cell showing the average variance ratio and its jackknife standard error.

\begin{table}
\centering\small
\begin{tabular}{ccccccc}
\toprule
& \multicolumn{3}{c}{Full Bootstrap} & \multicolumn{3}{c}{Reduced-Split Bootstrap} \\
\cmidrule(lr){2-4} \cmidrule(lr){5-7}
($n$, $\alpha$, $\beta$) & $\mathbf{t}_5$ & $\mathbf{t}_{10}$ & $\mathbf{t}_{16}$ & $\mathbf{t}_5$ & $\mathbf{t}_{10}$ & $\mathbf{t}_{16}$ \\
\midrule
(250, 0.6, 0) & \shortstack{1.038 \\ (0.077)} & \shortstack{0.976 \\ (0.067)} & \shortstack{0.965 \\ (0.064)} & \shortstack{1.035 \\ (0.075)} & \shortstack{0.960 \\ (0.065)} & \shortstack{0.957 \\ (0.064)} \\
\addlinespace
(500, 0.6, 0) & \shortstack{1.113 \\ (0.069)} & \shortstack{1.092 \\ (0.073)} & \shortstack{1.170 \\ (0.081)} & \shortstack{1.139 \\ (0.071)} & \shortstack{1.101 \\ (0.074)} & \shortstack{1.158 \\ (0.081)}  \\
\addlinespace
(1000, 0.6, 0) & \shortstack{1.128 \\ (0.069)} & \shortstack{1.096 \\ (0.069)} & \shortstack{1.057 \\ (0.072)} & \shortstack{1.120 \\ (0.070)} & \shortstack{1.094 \\ (0.070)} & \shortstack{1.040 \\ (0.071)}  \\
\addlinespace
(250, 0.6, 0.75) & \shortstack{1.028 \\ (0.063)} & \shortstack{1.013 \\ (0.064)} & \shortstack{1.031 \\ (0.065)} & \shortstack{1.036 \\ (0.064)} & \shortstack{1.011 \\ (0.065)} & \shortstack{1.037 \\ (0.068)}  \\
\addlinespace
(500, 0.6, 0.75) & \shortstack{1.043 \\ (0.075)} & \shortstack{1.028 \\ (0.070)} & \shortstack{1.066 \\ (0.064)} & \shortstack{1.034 \\ (0.075)} & \shortstack{1.019 \\ (0.070)} & \shortstack{1.062 \\ (0.065)}  \\
\addlinespace
(1000, 0.6, 0.75) & \shortstack{0.952 \\ (0.067)} & \shortstack{0.976 \\ (0.064)} & \shortstack{1.040 \\ (0.073)} & \shortstack{0.952 \\ (0.067)} & \shortstack{0.969 \\ (0.063)} & \shortstack{1.030 \\ (0.071)}  \\
\addlinespace
(250, 1.2, 0) & \shortstack{1.227 \\ (0.073)} & \shortstack{1.168 \\ (0.073)} & \shortstack{1.094 \\ (0.072)} & \shortstack{1.218 \\ (0.073)} & \shortstack{1.162 \\ (0.073)} & \shortstack{1.095 \\ (0.072)}  \\
\addlinespace
(500, 1.2, 0) & \shortstack{0.958 \\ (0.062)} & \shortstack{1.023 \\ (0.069)} & \shortstack{1.018 \\ (0.067)} & \shortstack{0.968 \\ (0.064)} & \shortstack{1.017 \\ (0.070)} & \shortstack{1.022 \\ (0.068)}  \\
\addlinespace
(1000, 1.2, 0) & \shortstack{1.062 \\ (0.062)} & \shortstack{1.052 \\ (0.064)} & \shortstack{1.035 \\ (0.064)} & \shortstack{1.061 \\ (0.063)} & \shortstack{1.050 \\ (0.065)} & \shortstack{1.041 \\ (0.065)}  \\
\addlinespace
(250, 1.2, 0.75) & \shortstack{1.169 \\ (0.078)} & \shortstack{1.143 \\ (0.073)} & \shortstack{1.139 \\ (0.069)} & \shortstack{1.152 \\ (0.080)} & \shortstack{1.120 \\ (0.075)} & \shortstack{1.121 \\ (0.071)}  \\
\addlinespace
(500, 1.2, 0.75) & \shortstack{1.100 \\ (0.070)} & \shortstack{1.037 \\ (0.069)} & \shortstack{1.032 \\ (0.070)} & \shortstack{1.096 \\ (0.071)} & \shortstack{1.041 \\ (0.070)} & \shortstack{1.030 \\ (0.071)}  \\
\addlinespace
(1000, 1.2, 0.75) & \shortstack{0.980 \\ (0.064)} & \shortstack{0.993 \\ (0.068)} & \shortstack{0.933 \\ (0.060)} & \shortstack{0.996 \\ (0.066)} & \shortstack{0.998 \\ (0.068)} & \shortstack{0.936 \\ (0.061)}  \\
\bottomrule
\end{tabular}
\caption{Variance ratios (with jackknife standard errors) comparing the two different bootstrap-based variance estimates to the Monte Carlo variance for various $(n,\alpha,\beta)$.}
\label{tab:var_ratios}
\end{table}

Across all configurations, the variance ratios are generally close to one: most fall within one jackknife standard error of $\text{Ratio}_j = 1$, and nearly all lie within two standard errors. Notably, the reduced-split bootstrap yields results nearly indistinguishable from the full bootstrap, despite using far fewer resamples. Given this substantial efficiency gain, we recommend the reduced-split approach for practical use.

%\FloatBarrier

\subsection{Optimal Estimation}\label{sec:WLSoptimal}

The tail index estimator studied in this paper is based on weighted least squares (WLS) and depends on a user-specified probability vector $\mathbf{t}_k$. This naturally raises the question of how to choose $k$ in practice. While a full treatment of this selection problem is beyond the scope of this paper, we explore initial strategies that may guide future work. \textcolor{black}{We consider this question here after having established the performance of the WLS estimator and the reduced-split bootstrap, since the selection and combination strategies developed below rely directly on both sets of results.}

The results presented here should not be taken as the final word on optimal estimation in this context. Rather, our goal is to demonstrate how the split-sample approach can be further improved by leveraging the bootstrap methodology developed for variance estimation in Section~\ref{sec:var_est}. In doing so, we provide a constructive proof of concept: tools already developed for variance estimation can be repurposed to aid in estimator selection and combination. In particular, this approach helps mitigate sensitivity to the specific choice of $\mathbf{t}_k$ in the WLS procedure. To that end, we revisit the $M = 500$ simulation replications and, for each dataset, compute three WLS estimators corresponding to equally spaced quantile vectors $\mathbf{t}_k$ with $k \in \{5, 10, 16\}$. As a performance benchmark, we use the WLS estimator that achieves the lowest empirical mean squared error (MSE) for a given configuration as the \textit {oracle}.

In addition to this idealized benchmark, we consider four practical strategies:
(i) random selection among the three WLS estimators;
(ii) an equal-weighted average of the three;
(iii) an optimal linear combination based on the inverse of the reduced-split bootstrap covariance estimate $\widetilde{\bm{\Omega}}_{250}$; and
(iv) selection of the single estimator with the lowest estimated variance under $\widetilde{\bm{\Omega}}_{250}$.

Although the oracle is useful for comparison, it is defined in hindsight. As it requires knowledge of which estimator performs best across many repeated samples, it is not a practical strategy. The remaining approaches offer varying degrees of adaptivity and computational burden.  The random selection strategy is included primarily as a cautionary example of a ``worst-case'' approach. 

%For each method, we report both $n^{1/2} \, \text{SD}$ and $n^{1/2} \, \text{RMSE}$ across the $M = 500$ replications, see Table~\ref{tab:selection_SD}. The $n^{1/2}$ scaling not only improves the readability of results but also reflects the asymptotic behavior of the estimators and facilitating comparison across sample sizes.

%\textcolor{black}{New we need to choose one of these tables and/or move on to the appendix, I favor keeping MSE ratios and perhaps putting the other one in the appendix?} 

The simulation results are presented in Table~\ref{tab:MSE_ratios} as MSE ratios. That is, each entry reports the MSE of a given strategy (Random, Avg, OptLin, or MinVar) divided by the oracle MSE. By reporting MSE ratios rather than absolute values, we normalize across sample sizes and focus on relative efficiency. Values below one indicate that a method, on average, outperforms the best single WLS estimator in hindsight; values above one reflect a loss in efficiency. This perspective highlights not only which strategies approach the oracle benchmark, but also whether any combination rule can, in practice, match or improve upon it in terms of MSE.

\begin{table}
\centering\small
\begin{tabular}{cccccc}
\toprule
$(\alpha,\beta)$ & $n$ & Random & Avg & OptLin & MinVar\\
\midrule
(0.6, 0) & 250 & 1.261 & 1.148 & 1.505 & 1.380\\
(0.6, 0) & 500 & 1.142 & 1.024 & 1.154 & 1.208\\
(0.6, 0) & 1000 & 1.160 & 0.999 & 1.089 & 1.193\\
\addlinespace
(0.6, 0.75) & 250 & 1.241 & 1.112 & 1.142 & 0.982\\
(0.6, 0.75) & 500 & 1.136 & 1.024 & 1.030 & 1.045\\
(0.6, 0.75) & 1000 & 1.032 & 0.981 & 1.044 & 1.026\\
\addlinespace
(1.2, 0) & 250 & 1.097 & 1.021 & 1.117 & 1.019\\
(1.2, 0) & 500 & 1.080 & 1.042 & 1.055 & 1.023\\
(1.2, 0) & 1000 & 1.016 & 0.922 & 0.976 & 0.987\\
\addlinespace
(1.2, 0.75) & 250 & 1.117 & 1.095 & 1.082 & 1.089\\
(1.2, 0.75) & 500 & 1.099 & 1.025 & 1.004 & 1.029\\
(1.2, 0.75) & 1000 & 1.063 & 0.986 & 1.003 & 1.023\\
\bottomrule
\end{tabular}
\caption{MSE ratios relative to the ``oracle’’ estimator for selection strategies: random choice (Random), equal-weight average (Avg), optimal linear combination (OptLin), and minimum-variance selector (MinVar).}
\label{tab:MSE_ratios}
\end{table}

Inspection of Table~\ref{tab:MSE_ratios} reveals that the random selection strategy performs the worst in most settings, often yielding substantially higher MSEs than the alternatives. This confirms its role as a cautionary baseline rather than a viable strategy. Among the more structured approaches, performance differences are more nuanced. No single method dominates across all settings, with each outperforming the others in at least one configuration. Selection is particularly challenging when the sample size is small ($n = 250$), and the distribution is heavy-tailed ($\alpha = 0.6$), where individual estimators may be biased or unstable.

As the sample size increases to $n = 1000$, all three structured strategies exhibit strong performance, frequently achieving MSE ratios at or below one. This suggests that combining information across estimators can yield improvements over any single WLS estimator. Among the methods, the equal-weight average stands out for its consistent performance and for attaining MSE ratios below one in all large-sample settings. Its robustness likely reflects the fact that it avoids estimating the covariance structure, thereby sidestepping additional sources of variability.

Finally, it is important to note that these results are based on only three candidate WLS estimators. A more comprehensive evaluation is needed to assess selection strategies over larger candidate sets, especially when relying on bootstrap-based covariance estimates.

%\FloatBarrier

\section{Real-Data Illustration: XRP Returns}\label{sec: Real-Data}

To illustrate the split-sample estimation framework, we analyze daily log-returns of XRP (daily closing prices) from August~7,~2013, to May~27,~2025, comprising $T=4281$ observations. XRP, also called Ripple, designed for fast, low-cost cross-border payments and, at the time of writing, ranked fourth by 24-hour trading volume (following Tether, Bitcoin, and Ethereum). Log-returns are defined as $R_t = \log(P_{t+1}) - \log(P_t)$, where $P_t$ denotes the USD closing price on day $t=1,\ldots,T-1$.

The use and public perception of cryptocurrencies has evolved substantially over the past decade, with major changes in regulation, adoption, and trading behavior. Accordingly, we do not treat the full sample as homogeneous over time. Instead, we divide the data into three approximately equal segments and focus on the earliest and most recent periods, each containing $n = 1437$ log-return observations.

\begin{figure}
\centering
\includegraphics[width=0.95\linewidth]{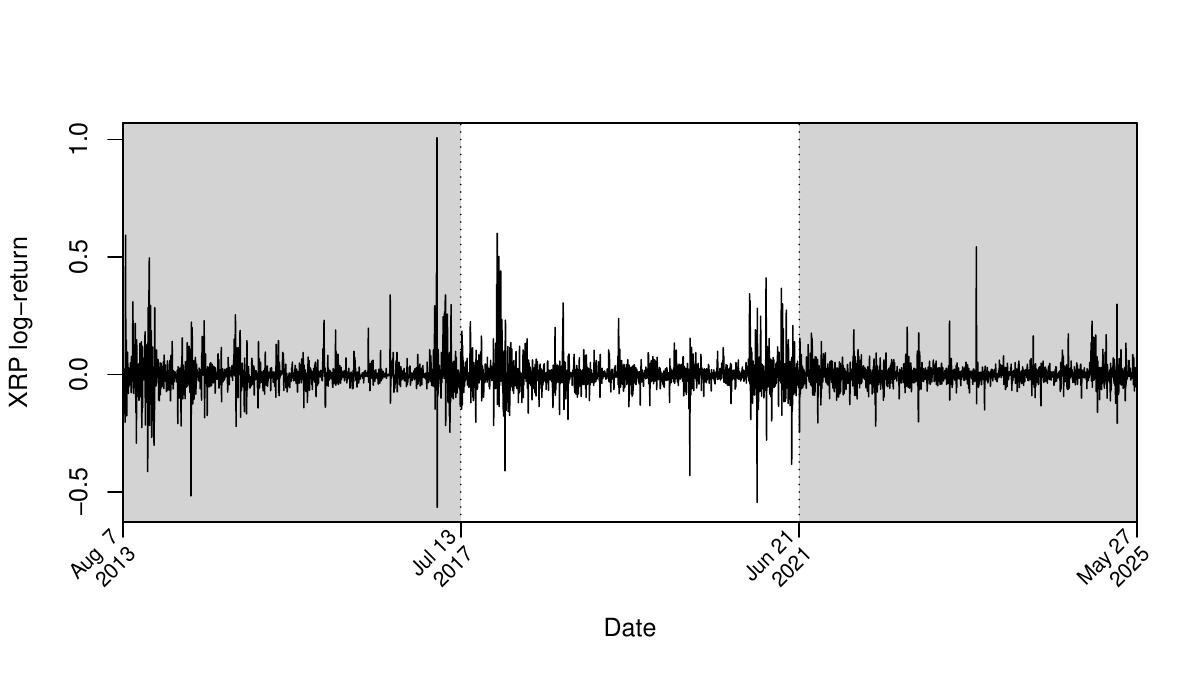}
\caption{Time series of daily XRP log-returns from August 7, 2013 to May 27, 2025. Shaded regions represent the early and recent periods used in analysis.}
\label{fig:xrp_line}
\end{figure}

The methodology assumes independent observations. To empirically assess this, we computed the rank-based lag-1 autocorrelation, obtaining $r_{1} = -0.02$, which is not statistically significant at the 5\% level. This provides some justification for treating observations within each period as approximately independent. Figure~\ref{fig:xrp_line} displays the XRP log-returns over time, with vertical dashed lines indicating the period splits and shaded regions highlighting the early and recent intervals used for analysis.

\begin{figure}
\centering
\includegraphics[width=0.99\linewidth]{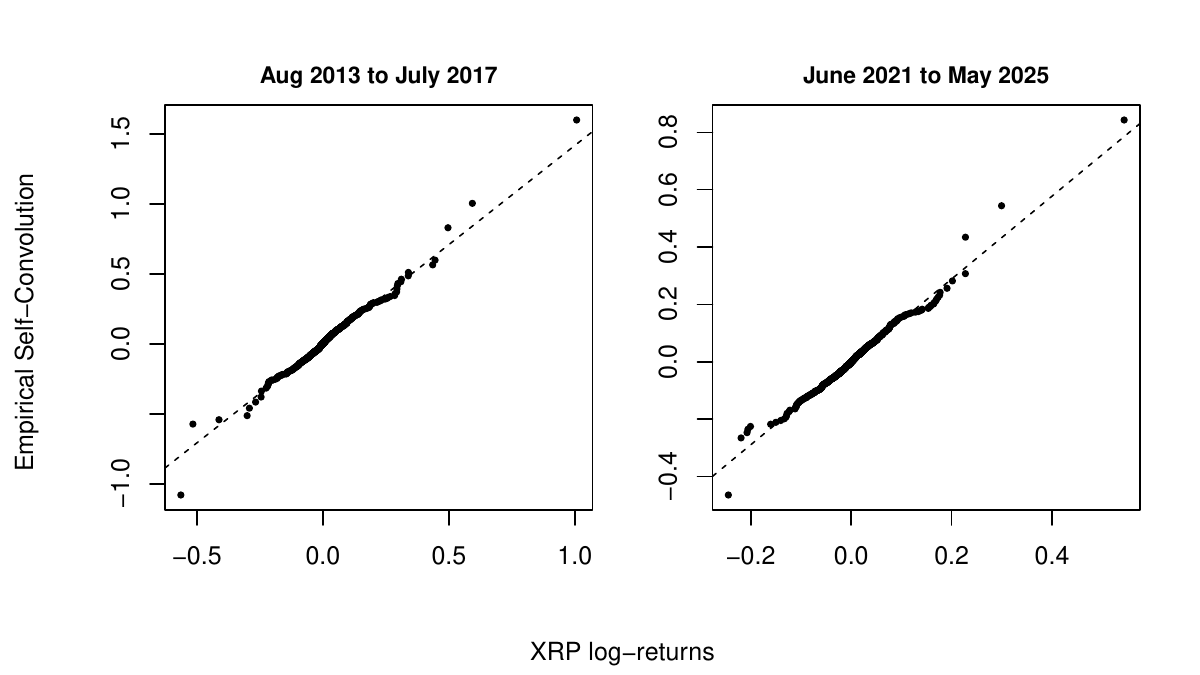}
\caption{QQ-plots of XRP log-returns versus their empirical self-convolutions in the early and recent periods. Dashed lines show quantile-matched linear fits.}
\label{fig:xrp_qq}
\end{figure}

Figure~\ref{fig:xrp_qq} presents quantile-quantile (QQ) plots comparing the empirical distribution of log-returns to their self-convolutions (pairwise sums) for the two periods being considered. Each plot includes a quantile-matched linear fit. The approximate linearity of the plots provides support for the use of a location-scale model, while the differing scales between periods reflect temporal changes in return magnitudes.

We applied the proposed split-sample estimator using $B = 500$ random splits with equal assignment probability $p = 0.5$. Three weighted least squares (WLS) estimators were computed using equally spaced quantile vectors of lengths $k = 5$, $10$, and $16$. Additionally, we considered an interquartile range (IQR) ratio estimator, defined as $\mathrm{IQR}_y / \mathrm{IQR}_x$, where the subscripts $y$ and $x$ correspond to the convolution and baseline samples, respectively. Since the IQR and WLS methods estimate the tail index on the $\sigma$ scale, we transformed these according to $\alpha = \log(2)/\log(\sigma)$, applied after averaging across the $B$ splits.

To quantify uncertainty, we employed the reduced-split bootstrap with $n_{\text{boot}} = 500$ resamples and $B_{\max} = 25$ splits per resample, extrapolating to $B = 500$ using the procedure described in Section~\ref{sec:bootstrap_extrap} to estimate the standard errors of the respective estimators.

We further considered two optimally weighted linear combinations using the bootstrap covariance matrix. The first (Opt A) combined all four split-sample estimators, including the IQR ratio; the second (Opt B) used only the three WLS estimators. Reported standard errors for these combinations reflect the variance of the weighted combination but do not account for variability in the estimated weights; therefore, they may underestimate the true uncertainty. A more rigorous nested bootstrap could address this, but is beyond our current scope.

For comparison, we include two classical estimators of $\alpha$: the Koutrouvelis and McCulloch quantile estimators, with standard errors estimated via 500 bootstrap resamples. We also report maximum likelihood estimates (MLE) from the \texttt{StableEstim} package in \texttt{R}, with standard errors from the observed information matrix (bootstrap standard errors were not computed due to computational demands).

\begin{table}
\centering\small
\begin{tabular}{lcc}
  \toprule
$\alpha$ Estimator & Aug '13 to July '17 & June '21 to May '25 \\ 
  \hline
Koutrovelis  & 1.1751 (0.0554) & 1.5230 (0.0426) \\ 
McCulloch & 1.2558 (0.0460) & 1.5871 (0.0420) \\ 
IQR Ratio & 1.0189 (0.0449) & 1.3270 (0.0630) \\ 
$\mathbf{t}_{5}$ & 1.1875 (0.0355) & 1.4692 (0.0424) \\ 
$\mathbf{t}_{10}$ & 1.2460 (0.0376) & 1.4830 (0.0405) \\ 
$\mathbf{t}_{16}$ & 1.2534 (0.0355) & 1.4835 (0.0385) \\ 
Opt A & 1.1673 (0.0301) & 1.4532 (0.0355) \\ 
Opt B & 1.2215 (0.0326) & 1.4793 (0.0370) \\ 
MLE & 1.1502 (0.0336) & 1.4718 (0.0402) \\ 
   \bottomrule
\end{tabular}
\caption{Estimates of the stability index $\alpha$ with standard errors (in parentheses) for the early and recent XRP periods. All standard errors are bootstrap-based, except MLE, which uses the observed information matrix.}
\label{tab:xrp_alpha}
\end{table}

Table~\ref{tab:xrp_alpha} summarizes the estimated stability indices and standard errors. Within each period, estimates are reasonably consistent across methods. For 2013–2017, estimates range from 1.0189 (IQR Ratio) to 1.2558 (McCulloch), with MLE and Opt A near the center. For 2021–2025, estimates range from 1.3270 to 1.5871, again with MLE and Opt A centrally located. The observed increase in $\alpha$ over time suggests a shift toward thinner tails, potentially reflecting greater market maturity and institutional participation. Nonetheless, $\alpha$ remains below $2$, consistent with infinite variance behavior.

In terms of standard error, the WLS estimators generally outperform the classical methods, particularly in the early period. In the most recent period, WLS estimators again yield smaller standard errors than Koutrouvelis, and all but one (WLS with $k=5$) outperform McCulloch. Although inefficient individually, the IQR ratio enhances performance when combined: Opt A (including IQR) achieves smaller standard errors than Opt B (excluding IQR). Additional analyses (not shown) indicate that replacing the IQR with an additional WLS estimator does not yield comparable gains, suggesting that the IQR ratio provides orthogonal information.

The optimal linear combinations attain the smallest standard errors in both periods, even outperforming MLE. However, two caveats apply: (i) standard errors for Opt A and Opt B do not reflect weight estimation uncertainty, and (ii) MLE standard errors are derived from the information matrix, not the bootstrap, complicating direct comparisons. These findings highlight the promise of the split-sample approach while motivating further work on post-selection inference and more refined variance estimation.

\section{Conclusion}\label{sec: Conclusion}

This paper introduced a split-sample estimator for the stability index $\alpha$, leveraging the two-sum closure property within a semiparametric, quantile-based framework. This formulation is closely connected to the class of stable distributions. By repeatedly partitioning a single sample and combining location–scale weighted least squares (WLS) estimators, the method achieves competitive root mean squared error (RMSE) performance compared with non-maximum-likelihood (MLE) alternatives for the tail index. The reduced-split bootstrap further enables fast and accurate variance estimation without the computational burden of full resampling. This is an important advantage given the time-consuming nature of existing MLE implementations in large samples.

In summary, the key contributions are as follows: a simple, likelihood-free estimator for the stability index $\alpha$ based solely on order statistics; variance reduction through repeated splitting and aggregation; an efficient bootstrap procedure using covariance extrapolation to avoid computationally intensive full resampling; and competitive empirical performance relative to widely used tail index estimators.

Several directions merit future work. \textcolor{black}{First, while we have used equispaced quantile levels for the WLS estimator and observed strong overall performance, adaptive selection of these levels may further enhance the method's efficiency. Such an ``optimal'' choice plausibly depends on features of the underlying distribution, such as skewness and tail heaviness. A systematic exploration of such distribution-dependent selection is beyond the scope of the present work.} Related, alternative two-sample estimation techniques could be adapted to the split-sample framework. Second, addressing post-selection inference remains important, particularly for estimator aggregation. Third, extending the estimator combination framework to include non-split estimators could unlock additional efficiency gains. Finally, further characterizing the broader class of distributions satisfying two-sum closure is an interesting open question.

Overall, the proposed approach combines practical ease of use with strong statistical performance, offering a flexible and computationally attractive tool for tail index estimation in heavy-tailed settings.

\section*{Code and Data Availability}

The code for the implementation of the methods proposed in the article, as well as the data used to illustrate the methods, are available at \url{https://github.com/cpotgieter/sample-splitting-stable-alpha}. The repository contains the implementation of the proposed split-sample estimators as well as the reduced-split bootstrap procedure. Additional simulation results referenced in the manuscript, namely RMSE summaries comparing estimators on the $\sigma$ and $\alpha$ scales, are also provided in the repository. 

\bibliography{bibliography.bib}

\appendix

\section{Consistent Estimation of Distributional Functions} \label{sec:Appendix}

This appendix analyzes the sampling properties of the empirical distribution functions introduced in Section~\ref{sec:Spliting}, focusing on consistency and variance behavior under the random-splitting scheme.

\subsection*{Probability of Degenerate Splits}

We begin by noting that the probability of a degenerate split, where all observations fall into either the baseline or convolution group, is negligible in practice. Specifically, the probability that either empirical distribution function is undefined due to an empty group exhibits exponential decay, up to a polynomial factor,
\begin{align}
    \mathbb{P}(n_1=0 \,\cup\, n_2 = 0) &= \mathbb{P}(n_1=0) + \mathbb{P}(n_1 \geq n-1)\notag\\ 
    &= (1-p)^n + p^{n-1}\big\{n(1-p)+p\big\}\notag\\
    &= \text{O}\bigl\{n \max(p,1-p)^{n}\bigr\}.
\end{align}

\subsection*{Behavior of the Counting Measures}

Recall the baseline counting measure $\hat{N}_1(x)$ defined in \eqref{eq:counting_proc} associated with the baseline sample $\mathcal{X} = \{Z_i : \delta_i = 1\}$. Since the Bernoulli assignment indicators $\delta_i$ are independent of the observed values $Z_i$, the expected count at threshold $x$ is
\begin{align}
    \mathbb{E}\bigl\{\hat{N}_1(x)\bigr\} = npF_1(x)
\end{align}
where $F_1$ denotes the baseline distribution function of the original population. The covariance at two thresholds $x_1$ and $x_2$ is
\begin{align}
    \mathbb{C}\mathrm{ov}\bigl\{\hat{N}_1(x_1),\hat{N}_1(x_2)\bigr\} = np\Bigl[F_1\bigl\{\min(x_1,x_2)\bigr\} - pF_1(x_1)F_1(x_2)\Bigr].
\end{align}
These expressions follow from standard arguments, relying on the independence between the $\delta_i$ and $Z_i$.

Next, recall from \eqref{eq:counting_proc} the convolution counting measure $\hat{N}_2(x)$ associated with the convolution sample $\mathcal{Y} = \{Z_i + Z_j : \delta_i = \delta_j = 0,\, i<j\}$. Its expectation at threshold $y$ is
\begin{align}
    \mathbb{E}\bigl\{\hat{N}_2(y)\bigr\} = \frac{n(n-1)}{2}\,(1-p)^2\,F_2(y),
\end{align}
where $F_2(y)=\mathbb{P}(Z+Z' \le y)$ denotes the distribution function of the sum of two independent draws from the baseline distribution. The covariance at two thresholds $y_1$ and $y_2$ decomposes into contributions from identical and overlapping pairs,
\begin{align}
    &\mathbb{C}\mathrm{ov}\bigl\{\hat{N}_2(y_1),\hat{N}_2(y_2)\bigr\} 
    =\frac{n(n-1)}{2}\bigl\{(1-p)^2\,F_2(y_1) - (1-p)^4\,F_2(y_1)F_2(y_2)\bigr\}\notag\\[1mm]
    &\quad +\,n(n-1)(n-2)\bigr\{(1-p)^3\,H(y_1,y_2) - (1-p)^4\,F_2(y_1)F_2(y_2)\bigr\}
\end{align}
where
\begin{align}
    H(y_1,y_2)=\mathbb{P}(Z+Z' \le y_1,\; Z+Z'' \le y_2) \label{eq:definition_of_H}
\end{align}
for $Z, Z', Z''$ independently drawn from the baseline distribution.

\subsection*{Behavior of the Empirical Distribution Functions}

The properties of the empirical counting measures are utilized to derive results for the split-sample empirical distribution functions. We begin by analyzing the asymptotic behavior of the baseline sample EDF, $\hat{F}_1(x)$. Conditional on observing $n_1 = k \ge 1$ baseline observations, the mean and covariance are 
\begin{align}
    \mathbb{E}\bigl\{\hat{F}_1(x)\mid n_1=k\bigr\}=F_1(x)
\end{align}
and
\begin{align}
    \mathbb{C}\mathrm{ov}\bigl\{\hat{F}_1(x_1),\hat{F}_1(x_2)\mid n_1=k\bigr\}=\frac{F_1\bigl\{\min(x_1,x_2)\bigr\}-F_1(x_1)F_1(x_2)}{k},
\end{align}
which mirror the classic results for empirical CDFs under a deterministic sample size. Using the fact that $n_1$ is Binomial$(n,p)$, the unconditional mean is
\begin{align}
    \mathbb{E}\bigl\{\hat{F}_1(x)\bigr\} = F_1(x)\mathbb{P}(n_1\ge1)=F_1(x)\bigl\{1-(1-p)^n\bigr\}=F_1(x) + \mathrm{O}\bigl\{(1-p)^n\bigr\}.
\end{align}

Applying the law of total covariance gives
\begin{align}
    \mathbb{C}\mathrm{ov}\bigl\{\hat{F}_1(x_1),\hat{F}_1(x_2)\bigr\}
    &=\mathbb{E}\Bigl[\mathbb{C}\mathrm{ov}\bigl\{\hat{F}_1(x_1),\hat{F}_1(x_2)\mid n_1\bigr\}\Bigr]\notag\\[1mm]
    &\qquad+\mathbb{C}\mathrm{ov}\Bigl[\mathbb{E}\bigl\{\hat{F}_1(x_1)\mid n_1\bigr\},\mathbb{E}\bigl\{\hat{F}_1(x_2)\mid n_1\bigr\}\Bigr]\notag\\[1mm]
    &=\mathbb{E}\biggl[\frac{F_1\bigl\{\min(x_1,x_2)\bigr\}-F_1(x_1)F_1(x_2)}{n_1}\,\mathbb{I}(n_1>0)\biggr]\notag\\[1mm]
    &=\Bigl[{F_1\bigl\{\min(x_1,x_2)\bigr\}-F_1(x_1)F_1(x_2)}\Bigr]\mathbb{E}\biggl\{\frac{\mathbb{I}(n_1>0)}{n_1}\biggr\}.
\end{align}
We furthermore have
\begin{align}
    \mathbb{E}\left\{\frac{\mathbb{I}(n_1>0)}{n_1}\right\} = \frac{1}{np} + \mathrm{O}\big(n^{-2}\big),
\end{align}
and hence,
\begin{align}
    \mathbb{C}\mathrm{ov}\bigl\{\hat{F}_1(x_1),\hat{F}_1(x_2)\bigr\} = \frac{F_1\bigl\{\min(x_1,x_2)\bigr\}-F_1(x_1)F_1(x_2)}{np} + \mathrm{O}\big(n^{-2}\big).
\end{align}

Next, consider the convolution EDF $\hat{F}_2(y)$ as per \eqref{eq:split_edf_def}. Similar to the derivation for the baseline EDF. Conditional on $n_2 = k \geq 1$, we have that 
\begin{equation}
\mathbb{E}\bigl\{\hat{F}_2(y)\,|\, n_2 = k\bigr\} = F_2(y)
\end{equation}
and
\begin{equation}
\mathbb{C}\mathrm{ov}\bigl\{\hat{F}_2(y_1), \hat{F}_2(y_2) \mid n_2 = k\bigr\}
= \frac{\mathbb{C}\mathrm{ov}\bigl\{\hat{N}_2(y_1), \hat{N}_2(y_2)\bigr\}}{k^2}.
\end{equation}
Since $\mathbb{P}(n_2 = 0) = \mathbb{P}(n_1 \geq n-1) = np^{n-1}(1-p) + p^n$, it follows that 
\begin{equation}
\mathbb{E}\bigl\{\hat{F}_2(y)\bigr\} = F_2(y) + \mathrm{O}\bigl\{(1-p)^n\bigr\}.
\end{equation}

From the earlier derivation, we know the covariance of the convolution counts can be written as
\begin{align}
&\mathbb{C}\mathrm{ov}\bigl\{\hat{N}_2(y_1),\hat{N}_2(y_2)\bigr\} \notag \\
&\quad= n(n-1)(n-2) \bigl\{(1-p)^3 H(y_1,y_2) - (1-p)^4 F_2(y_1)F_2(y_2)\bigr\} + \mathrm{O}\big(n^2\big),
\end{align}
where $H(y_1, y_2)$ is defined in \eqref{eq:definition_of_H}. Hence, the unconditional covariance of $\hat{F}_2(y_1)$ and $\hat{F}_2(y_2)$, conditional on $n_2 > 0$, becomes
\begin{align}
\mathbb{C}\mathrm{ov}\bigl\{\hat{F}_2(y_1), \hat{F}_2(y_2)\bigr\}
= \mathbb{E}\Bigg[\frac{1}{n_2^2} \mathbb{C}\mathrm{ov}\bigl\{\hat{N}_2(y_1), \hat{N}_2(y_2)\bigr\}\mathbb{I}(n_2 > 0)\Bigg].
\end{align}
Using the approximation $n_2 = n(n-1)(1-p)^2 / 2 + \mathrm{o}\big(n^2\big)$, which reflects the expected number of convolution pairs under random assignment, we obtain the reciprocal squared scaling factor
\begin{align}
\frac{1}{n_2^2} = \frac{4}{n^2(n-1)^2 (1 - p)^4} + \mathrm{O}\big(n^{-3}\big),
\end{align}
which leads to the following large-sample approximation,
\begin{align}
&\mathbb{C}\mathrm{ov}\bigl\{\hat{F}_2(y_1), \hat{F}_2(y_2)\bigr\}\notag\\
&\quad= \frac{4(n-2)}{n(n-1)(1 - p)} \bigl\{H(y_1, y_2) - F_2(y_1)F_2(y_2)\bigr\} + \mathrm{O}\big(n^{-2}\big).
\end{align}

\end{document}